%% file: arXiv_v2.tex
\documentclass[11pt,a4,twoside]{article}

\usepackage{./styling/style-main}
\graphicspath{{Plots/}}
\usepackage{framed}
\usepackage{amsmath}

\newcommand{\bracket}[1]{\left\langle #1 \right\rangle}

 \usepackage[compat=1.0.0]{tikz-feynman}
 \usepackage{transparent}

\newcommand{\abs}[1]{\vert #1 \vert}
\newcommand{\bU}{\overline{U}}
\newcommand{\bV}{\overline{V}}

\newcommand{\SLO}{S_\text{LO}}
\newcommand{\SNLO}{S_\text{NLO}}

\newcommand{\bLO}{\phi_\text{LO}}
\newcommand{\bNLO}{\phi_\text{NLO}}

\newcommand{\ak}{\abs{\kappa}}

\title{Bubble Nucleation to All Orders}

\author{Andreas~Ekstedt\thanks{andreas.ekstedt@desy.de}\textsuperscript{~~,\,a,\,b\,c}}
\affil{a:~Department of Physics and Astronomy, Uppsala University, P.O. Box 256, SE-751 05 Uppsala, Sweden}
\affil{b:~II. Institute of Theoretical Physics, Universität Hamburg, D-22761, Hamburg, Germany}
\affil{c:~Deutsches Elektronen-Synchrotron DESY, Notkestr. 85, 22607 Hamburg, Germany}

\date{\today}
\begin{document}
	\maketitle
		\thispagestyle{plain}

	\begin{abstract}
This paper extends classical results by Langer and Kramers \cite{Langer:1969bc,KRAMERS1940284,PhysRevA.8.3230} and combines them with modern methods from high-temperature field theory\cite{Bodeker:1999ey,Arnold:1998cy,Blaizot:1993zk,Moore:2000jw,Moore:1998swa}. Assuming Langevin dynamics, the end-product is an all-orders description of bubble-nucleation at high temperatures. Specifically, it is shown that equilibrium and non-equilibrium effects factorize to all orders\te the nucleation rate splits into a statistical and a dynamical prefactor. The derivation clarifies, and incorporates, higher-order corrections from zero-modes\cite{Andreassen:2016cvx,Andreassen:2017rzq,GERVALS1976281}. The rate is also shown to be real to all orders in perturbation theory. The methods are applied to several models. As such, Feynman rules are given; the relevant power-counting is introduced; RG invariance is shown; the connection with the effective action is discussed, and an explicit construction of propagators in an inhomogeneous background is given.
The formalism applies to both phase and Sphaleron transitions. While mainly focused on field theory, the methods are applicable to finite-dimensional systems. Finally, as this paper assumes an effective Langevin description\cite{Blaizot:1993be,Blaizot:1993zk,Berera:2007qm,Gautier:2012vh,Moore:2000jw,Bodeker:1999ey,Arnold:1998cy}, all results only hold within this framework.
	\end{abstract}

	\input{./tex/introduction}

 \bibliographystyle{utphys}

{\linespread{0.6}\selectfont\bibliography{Bibliography}}

\end{document}

%% file: tex/introduction.tex
\section{Introduction}\label{sec:Introduction}
The observation of gravitational waves is a game-changer, granting us new eyes to gaze at the cosmos\cite{Abbott:2016blz,Hindmarsh:2020hop,Guo:2021qcq,Caprini:2019egz}. There is now a realistic chance of glimpsing phase transitions that occurred a few nanoseconds after the Big Bang: if the transition is first order. Contrary to continuous phase transitions, a first-order transition can leave tracks in the form of a stochastic gravitational-wave background. This is because these transitions proceed through nucleating bubbles \cite{Linde:1981zj,Langer:1969bc,PhysRevA.8.3230,Coleman:1980aw,Callan:1977pt,Coleman:1977py}. The interior of these bubbles is permeated by a true vacuum; the outside situated in a metastable vacuum. Once these bubbles expand they can generate gravitational waves through collisions and turbulence in the primordial plasma \cite{Caprini:2006jb,Hindmarsh:2017gnf,PhysRevD.92.123009,Hindmarsh:2013xza,Jinno:2016vai}. Furthermore, these bubbles might facilitate Electroweak Baryogenesis\te thus potentially explaining the observed Baryon asymmetry \cite{Kuzmin:1985mm,Shaposhnikov:1986jp,Shaposhnikov:1987tw,Cohen:1993nk}. 

Although, it is unclear if a first-order phase transition occurred in our cosmological history, not to mention whether gravitational-waves from such a transition can be detected by upcoming experiments~\cite{2017arXiv170200786A,Guo:2018npi,Kawamura:2006up,Harry:2006fi}. 

To answer these questions both model building \cite{Turok:1991uc,Athron:2019teq,Barger:2007im,Alves:2018jsw,Niemi:2021qvp,Bell:2020hnr,Niemi:2020hto,deVries:2017ncy,Hall:2019ank,Baldes:2018nel} and theoretical calculations \cite{Croon:2020cgk,Ekstedt:2021kyx,Farakos:1994kx,Kajantie:1995dw,Kajantie:1995kf,Rummukainen:1998nu,Gurtler:1997hr,Kajantie:1996mn,Gould:2021dzl,Dorsch:2018pat,Gould:2019qek} are required. Theoretical calculations, in particular, need to be improved to reduce uncertainties in the description of rate-of-production, and successive growth, of nucleating bubbles~\cite{Lofgren:2021ogg,Gould:2021oba,Baacke:1999sc,Garny:2012cg,Guo:2021qcq}.

 In the context of perturbative calculations, one issue is that calculations converge slowly at high temperatures, so higher-order corrections can be large. As such, calculating these higher-order corrections provide much-needed cross-checks. This is especially relevant for the nucleation rate, where radiative corrections can be enhanced for large bubbles.

To that end, this paper shows how to calculate the bubble-nucleation rate to higher orders. 

 The main result of this paper is that the nucleation rate factorizes as
\begin{align}\label{eq:NucleationRate}
	\Gamma=A_\text{dyn} \times A_\text{stat}
\end{align}
Here $A_\text{dyn}$ accounts for non-equilibrium effects and $ A_\text{stat}$ captures equilibrium effects. These factors are known as the dynamical and the statistical prefactor respectively, and they can be calculated order-by-order in perturbation theory.
Equation \ref{eq:NucleationRate} is derived in Section \ref{sec:NucleationRate}, and applied to a real-scalar model in Section \ref{sec:FieldTheory}. Crucially all derivations assume a classical Hamiltonian system. See \cite{Blaizot:1993be,Blaizot:1993zk,Berera:2007qm,Gautier:2012vh,Moore:2000jw,Bodeker:1999ey,Arnold:1998cy} for the details and applicability of this effective description.

It is important to stress that formulas similar to Equation \ref{eq:NucleationRate} have been suggested before\cite{Langer:1969bc,Linde:1981zj,PhysRevD.47.5304,Affleck:1980ac,Gould:2021ccf}. However, it is unclear from past literature how higher-order corrections should, even in principle, be included.

In this paper, the nucleation rate is derived from first principles, and the factorization in Equation \ref{eq:NucleationRate} is proved to all orders. Moreover, Feynman rules for calculating the statistical and dynamical prefactors are derived. It is also shown how to treat zero-modes at higher orders. In addition, explicit calculations show that the rate is renormalization-scale invariant to two-loops.

\section{Uncertainties for thermal escape}
When studying thermal escape in field theory, one looks for solutions to the classical equations of motion. Not any solution, rather, these \textit{bounce} solutions obey particular boundary conditions and are the field-theory equivalence of saddle points \cite{Coleman:1977py,Callan:1977pt,Coleman:1980aw}. The energy of the bounce solution\te denoted in three dimensions by $S_3$\te controls the probability for the system to transition from a metastable state to a lower-energy state. These thermal transitions result in nucleating bubbles.

An approximation for the nucleation rate at high temperatures is\cite{Linde:1981zj,Affleck:1980ac,PhysRevD.47.5304}
\begin{align}\label{eq:NaiveRate}
	\Gamma\approx A T^4 e^{-S_3/T}
\end{align}

Naively one expects the exponent to dominate. Then when the transition happens, around $S_3\approx 140 T $\cite{Guo:2021qcq,Caprini:2019egz}, the exponential prefactor should be sub-leading.

There are however a few subtleties with Equation \ref{eq:NaiveRate}. First, $S_3$ depends on the renormalization-scale, which introduces significant uncertainties \cite{Gould:2021oba}. Second, the prefactor can dominate the exponent \cite{Ekstedt:2021kyx,Croon:2020cgk,Gould:2021ccf}. Third, the form of Equation \ref{eq:NaiveRate} neglects non-equilibrium effects\cite{PhysRevA.8.3230,Langer:1969bc}.

These three problems\te in addition to gauge dependence \cite{Garny:2012cg,Lofgren:2021ogg,Hirvonen:2021zej}\te result in order-of-magnitude uncertainties for the rate. This is important as these uncertainties propagate to gravitational-wave predictions\cite{Guo:2021qcq,Gould:2021oba}. 

\subsection{High-temperature corrections}\label{sec:HighT}

Perturbative calculations converge slower at high-temperatures. To see this, consider a scalar field with mass $m^2$. At high temperatures this mass is changed by thermal corrections: to one loop $m^2\rightarrow m^2_\text{eff}=m^2 +a T^2$, where $a$ is a function of coupling constants. Now, at the transition one typically finds $T^2\gg m^2 \gg m^2_\text{eff}$. This means that two-loop thermal masses are of similar size as $m^2_\text{eff}$, and in addition, logarithms of the form $\log T^2/m^2_\text{eff}$ enhance higher-order corrections.

These problems can be removed by integrating out high-energy modes with momenta $ k\sim T$ \cite{Farakos:1994kx,Kajantie:1995dw,Croon:2020cgk}. The resulting effective theory lives in three spatial dimensions; effective masses and couplings in this theory automatically incorporate thermal resummations. This effective theory can be used to both calculate equilibrium quantities and the nucleation rate~\cite{Croon:2020cgk,Gould:2021ccf,Ekstedt:2021kyx,Hirvonen:2021zej}.
\subsection{Size of the exponential prefactor}
Another subtlety is that corrections to the bounce-action can be large.
To see this, it is useful to rewrite the rate as 
\begin{align}
-\log \Gamma \sim S_3/T-\log A.	
\end{align}
For large bubbles with radius $R$, the first term scales as $S_3 \sim R^2$ \cite{Coleman:1978ae,Coleman:1977py,Coleman:1980aw,PhysRevA.8.3230,PhysRevD.47.5304}. Meanwhile, the prefactor can be estimated as the one-loop difference in energy (the effective potential) between the true and metastable state. That is, 
\begin{align}
	\log A \sim R^3 \left[ V_\text{eff}(\phi_\text{true})-V_\text{eff}(\phi_\text{false})\right]\sim R^3 g^a,
\end{align}
for some $a$. We see that the calculation breaks down for $R g^a \sim  1$.

It is also possible for $A$ to be of the same order, or larger, than the exponent if there are heavy particles in the theory. For example, for a particle with mass $M$ the prefactor scales as $\log A\sim M^{3/2}$.
Therefore, if $M$ is large these particles should be integrated out before calculating the rate\cite{Lofgren:2021ogg,Gould:2021ccf,Ekstedt:2021kyx,Hirvonen:2021zej}. 
\subsection{Non-equilibrium dynamics }

Thermal escape is a non-equilibrium process. As such, damping and related effects are important \cite{Langer:1969bc,PhysRevD.47.5304,Linde:1981zj}. However, these effects are hidden in Equation \ref{eq:NaiveRate}. In addition, the saddle-point approximation is expected to break down for small damping\cite{RevModPhys.62.251}. This motivates going beyond leading-order results~\cite{Linde:1981zj,Affleck:1980ac}.

\subsection{Thermal escape and tunneling}
To see how Equation \ref{eq:NaiveRate} should be modified, it is informative to discuss the physics behind thermal escape.

The formulas for quantum tunneling and thermal escape look similar
\begin{align}\label{eq:Rate}
&	\Gamma_{T=0}=\left(\frac{S_B}{2\pi}\right)^2\left\vert \frac{\det([-\partial^2+V''(0)]) }{\det'[-\partial^2+V''(\phi_b)]}\right\vert^{1/2}e^{-S_B} \\&
\Gamma_{T\neq0}=\frac{\lambda}{2\pi}\left(\frac{S_3}{2\pi T}\right)^{3/2}\left\vert \frac{\det([-\partial^2+V''(0)]) }{\det'[-\partial^2+V''(\phi_b)]}\right\vert^{1/2}e^{-S_3/T}
\end{align}
However, the physics is quite different.

For tunneling, the rate is\footnote{Subtleties related to zero modes~\cite{Andreassen:2016cvx,Andreassen:2017rzq} are discussed in Section \ref{sec:ZeroModes}. }
\begin{align}
\Gamma_{T=0}=2\text{Im}e^{-S_\text{eff}}.	
\end{align}
The effective action is evaluated on a solution of $\delta S_\text{eff}[\phi_b]=0$. In essence the tunneling rate comes from a saddle-point approximation around the bounce solution. The functional determinant (one-loop contribution to $S_\text{eff}$) arises from fluctuations around the bounce, and higher-order terms in $S_\text{eff}$ consist of vacuum diagrams in the bounce background~\cite{Andreassen:2016cvx,Andreassen:2017rzq}. Crucially the effective action is imaginary due to a negative eigenvalue of $\delta^2 S_B[\phi_b]$.

The thermal-escape formula is similar to the tunneling decay rate. Indeed, except for the prefactor, the determinant and exponent could be the first terms of $e^{-S_\text{eff}/T}$, where $S_\text{eff}$ is the effective action in three dimensions. This has led to the conjecture~\cite{LINDE198137,Linde:1981zj}
\begin{align}
	\Gamma_{T \neq 0}=A_\text{dyn}\times 2 \text{Im} e^{-S_\text{eff}/T}.
\end{align}
The leading-order dynamical prefactor is $A_\text{dyn}=\frac{\lambda}{2\pi}$, and $\lambda$ is normally taken as $\sqrt{\abs{\kappa}}$, where $\kappa$ is the only negative eigenvalue of $\delta^2 S_3 [\phi_b]$.

Yet thermal escape is a classical process. 

To illustrate the physics it is useful to consider a classical particle moving in three dimensions. We assume that the potential of this particle has one metastable minimum at $\vec{x}=\vec{x}_S$, and a lower-energy minimum at $\vec{x}=\vec{x}_\text{TV}$. Furthermore, let us assume that these two minima are separated by a barrier. Lowest at $\vec{x}=\vec{x}_B$. Without loss of generality we take $\vec{x}_B$ to lie in the $x_1$ direction.

If this was a tunneling process, the particle would start at the metastable minimum and go through the barrier with some probability. The most likely path of escape, the bounce, results in a negative eigenvalue, which gives an imaginary energy. The rate is identified with $\Gamma =-2 \text{Im} \mathcal{E}$.

But the picture is different for thermal escape, where the particle travels over the barrier.
 That is, the probability for the particle to have an energy $E$ is proportional to $e^{-E/T}$. This means that the particle is most likely to escape where the barrier is lowest ($V_B=V(x)\vert_{x=\vec{x}_B}$), with probability $\Gamma\sim e^{-V_B/T}$. Although, the particle can also escape close-by $\vec{x}_B$ by moving slightly in the $x_2$ or $x_3$ directions.\footnote{This discussion assumes an overdamped system, and ignores kinetic contributions to the energy. See Section \ref{sec:GeneralDamping} for a discussion of the general case.}

 A better approximation is $\Gamma \sim \int_{x_1=x_B}d x_2 d x_3 e^{-V(x)/T}$. Crucially only variations in orthogonal directions to $\vec{x}_B$ are included in this integral. The rate is real because there is no integration in the concave $x_1$ direction.

What about non-equilibrium physics? When the particle jumps over the barrier, it first needs to get sufficient energy to escape, and then move to the other side. If the \textit{negative} curvature ($\kappa$) in the concave direction is large, the probability to get moving over the edge is naturally big. Likewise, large damping ($\eta$) slows down the particle. This is the physical reason for the leading-order dynamical prefactor: $A_\text{dyn}=\frac{\ak}{2\pi \eta}$.

Intuitively the Boltzmann factor describes the probability to jump up to, and the dynamical factor describes the flow from, the barrier. To leading order this flow is in the concave direction, while higher-order corrections make the flow veer down the barrier from orthogonal directions.

Note that it does not matter that $V$ is concave in the $x_1$ direction because we should only integrate in orthogonal directions to $x_1$ when calculating the rate. Thus no imaginary terms can appear in the saddle-point approximation.\footnote{This has been previously observed in \cite{Gould:2021ccf,RevModPhys.62.251,Berera:2019uyp}. However, care must be taken for generic damping. For example, away from the strict overdamping limit the integral should not be done at the saddle point. Yet the decay rate is still real for general damping coefficients, but the reason is more subtle. See Section \ref{sec:GeneralDamping}.}

 These considerations motivate the main result of this paper
 \begin{align}\label{eq:RateFormulaFirst}
 	\Gamma_{T\neq 0}=A_\text{dyn} e^{-\overline{S}_\text{eff}},
 \end{align} 
where the effective action omits contributions from negative eigenvalues. This means that the effective action is real to all orders.

Equation \ref{eq:RateFormulaFirst} is derived in the next section for an overdamped system; an equivalent derivation for general damping is given in Section \ref{sec:GeneralDamping}. The complete nucleation rate, in field theory, is given in Section \ref{sec:FinalFacRate}.

\section{The nucleation rate}\label{sec:NucleationRate}
In this section we derive a complete formula for the nucleation rate. To that end, consider a system with $n$ degrees of freedom, indexed by $i$. Translating the results to field theory is straightforward, and is done in Section \ref{sec:FieldTheory}. 

For simplicity this section assumes an overdamped system. While not generic, an overdamped system makes the physics transparent. The steps are similar for the general-damping case, differing only by longer intermediate formulas, which are given in Section \ref{sec:GeneralDamping}.

As mentioned, the calculation of the nucleation rate is classical.
This follows because field theory is, to first approximation, classical at high temperatures \cite{Bodeker:1999ey,Arnold:1998cy,Blaizot:1993zk,Moore:2000jw,Moore:1998swa}.  This is indicated by the Boltzmann factor
\begin{align}
	n(\omega)=\frac{1}{e^{\omega/T}-1}.
\end{align}

At high temperatures $T \gg \omega$, the occupation number is large: $n(\omega)\approx T/\omega \gg 1$. Equivalently, the system is classical if the thermal-wavelength is small: $\lambda_T^{-1} \lambda_\text{dB} \sim  \frac{T}{\omega}\gg \hb$.

However, although nucleation occurs at low energies, large energy modes with $\omega \sim T$ still contribute through loops. These high-energy modes can be integrated out, and give effective couplings; effective damping parameters; and effective thermal noise~\cite{Bodeker:1999ey}.

In this section the damping is taken as a free parameter. See~\cite{Bodeker:1999ey,Arnold:1998cy,Blaizot:1993zk,Moore:2000jw,Moore:1998swa} for calculations, and discussions, of the damping for specific models.

\subsection{Classical nucleation theory}
This section reviews classical nucleation theory. We stress that the results in this first subsection are known \cite{RevModPhys.62.251,Langer:1969bc,Berera:2019uyp,KRAMERS1940284,Farkas+1927+236+242}.

The Einstein summation convention is used for the indices $i,j,k,l$, where the vertical position of these indices is unimportant.

Consider the Fokker-Planck equation for a system with large damping (also known as  the Smoluchowski equation)\cite{Fokker,1915AnP}
 \begin{align}\label{eq:FokkerPlanck}
	\partial_t P=-\partial_i J^i,\quad J^i=-\left[\frac{1}{\eta} P \partial_i V+\frac{T}{\eta} \partial_i P\right].
\end{align}
Here $V(x)$ is a generic potential, $\partial^i$ is the derivative with respect to $x^i$, and $\eta$ is the damping.\footnote{ It is assumed that all components [$x^i$] have the same damping $\eta$.}

The function $P(x, t;x_0,t_0)$ is the probability to find the particle at a position $x$, given that it was at $x_0$ at time $t_0$. Then, if $P$ is initially localized to a metastable state we can ask what is the probability-flow from this state.

Determining $P$ is in general hard, yet to calculate the rate it suffices to determine $P$ close to the barrier. The rate can then be approximated by the probability flow, defined by $J^i$, across this barrier\cite{Langer:1969bc,RevModPhys.62.251,KRAMERS1940284}:
\begin{align}\label{eq:RateDef}
\Gamma=\frac{\int_\text{barrier} \vec{dS}\cdot \vec{J}}{\int_\text{metastable} d^nx P}.
\end{align}  

The calculations become tractable if one introduces artificial sources and sinks so that $P$ is static~\cite{Langer:1969bc,KRAMERS1940284,RevModPhys.62.251,Berera:2019uyp}. In that case the problem is reduced to solving the static Fokker-Planck equation. After the equation is solved, and we have our current, the sinks and sources can be removed. This is known as the flux-over-population method. Although, it should be mentioned that this method fails for underdamped systems\cite{RevModPhys.62.251}.

To use the flux-over-population method, consider first the metastable state, taken to be at $\vec{x}=\vec{x}_S$. The (static) equilibrium solution is $P_0=e^{-V(x)/T}$, which is the normal Boltzmann distribution and follows from Equation \ref{eq:FokkerPlanck}.  Further, if the potential is of the form $V(x+x_S)=V_S+\frac{1}{2}x_{i}\omega_S^{ij}x_j+\ldots$ close to the metastable point, the denominator in Equation \ref{eq:RateDef} can be calculated with a saddle-point approximation.

Next, the idea is to determine $P$ close to the barrier\te generally a saddle point. This requires two approximations. First, we assume that the surface integral in Equation \ref{eq:RateDef} is dominated by the saddle point. Second, we assume that the surface can be approximated as a surface normal to the saddle point.

To find $P$ close to the saddle point it is useful to follow Langer and make the ansatz $P=\zeta(\vec{x})  e^{-V(x)/T}$. The function $\zeta(\vec{x})$ describes deviations from thermal equilibrium and behaves as 
\begin{align}
	& \lim_{\vec{x}\rightarrow \vec{x}_S} \zeta(\vec{x})=1,
	\\& \lim_{\vec{x}\rightarrow \vec{x}_\text{TV}} \zeta(\vec{x})=0,
\end{align}
where $\vec{x}_\text{TV}$ denotes the true (lower-energy) vacuum state.

Using this ansatz in Equation \ref{eq:FokkerPlanck} one finds
\begin{align}\label{eq:ZetaEquation}
	T \partial_i \partial^i \zeta-\partial^i V\partial^i \zeta=0.
\end{align}

To solve this equation, assume that the saddle point is at $x_B^i$. At this point the potential is to first approximation quadratic: $V(x_B+x)=V_B+\frac{1}{2}x_i \omega^{ij}x_j$. Hence
\begin{align}
	T \partial_i \partial^i \zeta-x_i\omega^{ij}\partial^j \zeta=0.
\end{align}

This equation can be solved with the ansatz $\zeta(\vec{x})=\zeta(u)$, where $u$ is a linear combination of $x^i$. Namely, $u=\bU^i x_i$. Then
\begin{align}\label{eq:ZetaAnsatsWU}
	T \bU^2 \zeta''(u)-x_i \omega^{ij}\bU^j \zeta'(u)=0,
\end{align}
where $\bU^i$ must be an eigenvector of $\omega^{ij}$ for consistency. That is
\begin{align}
	\omega^{ij}\bU^j=\kappa \bU^i.
\end{align}

For simplicity we choose the normalization $\bU^2=1$. The solution to Equation \ref{eq:ZetaAnsatsWU} is\cite{Langer:1969bc,RevModPhys.62.251}
\begin{align}\label{eq:Zeta0}
	&\zeta(u)=\frac{1}{\sqrt{2\pi \abs{\alpha}}}\int_{u}^\infty du' e^{-u'^2/(2\abs{\alpha})}, \quad \alpha=\frac{T}{\kappa}.
\end{align}

The boundary conditions for $\zeta(u)$ force $\kappa <0$, and so $\bU^i$ is the eigenvector corresponding to a negative eigenvalue. As expected, deviations from equilibrium only depend on the negative-eigenvalue direction. 

Given $\zeta(u)$, the current is
\begin{align}
	J^i=-\left[\frac{1}{\eta}  \partial_i V+\frac{T}{\eta}\partial_i\right] P=-\bU^i \frac{T}{\eta} \zeta'(u)P_0,
\end{align}
where  $P_0=e^{-V(x)/T}$.

To find the rate we integrate the current on a surface normal to $u=0$. This gives the rate\footnote{Here, following \cite{RevModPhys.62.251},  the delta function has been rewritten as $\delta(u)=\frac{1}{2\pi}\int_{-\infty}^{\infty} dk  e^{i k u}$.}
\begin{align}\label{eq:LeadingOrderRate}
	\Gamma\propto\int_{u=0} d^n x \bU^i J^i=\frac{1}{\eta}\int d^nx \int_{-\infty}^\infty \frac{dk}{2\pi}\frac{1}{\sqrt{2\pi \abs{\alpha}}} e^{i k u-V(x)/T},
\end{align}
where to leading order $V(x+x_B)\approx V_B+\frac{1}{2}x_i \omega^{ij}x_j$. Doing the integrations one finds \cite{Langer:1969bc,RevModPhys.62.251}
\begin{align}\label{eq:RateOverdampingLO}
	\Gamma\propto\frac{\abs{\sqrt{\kappa}}}{2\pi \eta} \left[\bar{\det}' \omega\right]^{-1/2}e^{-\frac{V_B}{T}},
\end{align}
where $\bar{\det}'$ means that all negative and zero eigenvalues are excluded, see Section \ref{sec:ZeroModes} for the details. Crucially $u=0$ excludes the negative-eigenvalue direction, and the determinant is manifestly real.

Normalizing with the denominator in Equation \ref{eq:RateDef} gives the known result\cite{RevModPhys.62.251,Langer:1969bc,Berera:2019uyp}
\begin{align}\label{eq:LORate}
	\Gamma=\frac{\abs{\sqrt{\kappa}}}{2\pi \eta} \left[\frac{\bar{\det}' \omega}{\det \omega_S}\right]^{-1/2}e^{-\frac{(V_B-V_S)}{T}}.
\end{align}
We refer to this as the leading-order rate from now on.

Consider now corrections to Equation \ref{eq:LORate}, of which there are two kinds. First, corrections to the dynamical prefactor come from including more terms for $V$ in Equation \ref{eq:ZetaEquation}. Second, the statistical prefactor receives corrections from additional terms in $V$ via Equation \ref{eq:LeadingOrderRate}. To leading order we can write the rate as:\vspace*{-0.3cm}
\begin{align}\label{eq:LeadingOrderPrefactor}
	&\Gamma=A_\text{dyn}\times A_\text{stat},
	\\& A_\text{dyn}=\frac{\abs{\sqrt{\kappa}}}{2\pi \eta} \hspace{0.3cm} \&  \hspace{0.3cm}  A_\text{stat}= \left[\bar{\det}' \omega\right]^{-1/2}e^{-\frac{V_B}{T}}.
\end{align}

As the next two sections show, the statistical prefactor consists of normal (exponentiated) vacuum diagrams; the dynamical prefactor, in field language, consists of operator insertions.

Henceforth we leave the metastable normalization implicit. This normalization factor is straightforward to calculate via the effective potential \cite{Arnold:1992rz,PhysRevD.7.1888}.

\subsection{Corrections to the statistical prefactor}\label{sec:StatPrefactor}

Consider the statistical prefactor. As mentioned, the negative eigenvalue does not cause any issues because everything is calculated at $u=0$.

To proceed, consider the integral in Equation \ref{eq:LeadingOrderRate}: 
\begin{align}
	Z[ \bU]=	\int d^n x e^{i k u-1/T\frac{1}{2} \left(x_i \omega^{ij}x_j\right)}.
\end{align}
Integrating over $k$ enforces $u=0$, and we leave this integration for last.

Barring some subtleties with the negative eigenvalue, the integrals give
\begin{align}
Z[ \bU]=	\left[\text{det}' \omega/(2\pi) \right]^{-1/2}e^{-\frac{1}{2}k^2 T \bU^i\omega^{-1}_{ij} \bU^j}.
\end{align}
Technically $\bU^i\omega^{-1}_{ij} \bU^j<0$, so the integral over $k$  formally diverges. Being more careful one can let $k\rightarrow -i k$ and do a Wick-rotation. This gives
\begin{align}\label{eq:GeneratingFunc}
	Z[\bU]=\left[\abs{\text{det}' \omega/(2\pi)} \right]^{-1/2}e^{\frac{1}{2}k^2 T \bU^i\omega^{-1}_{ij} \bU^j}.
\end{align}
We identify $Z[\bU]$ as a generating function: correlators can be calculated via 
\begin{align}
	<x^i x^j>\equiv \int_{-\infty}^\infty d k \int d^n x x^i x^j e^{i k u-1/T \left(x_i \omega^{ij}x_j\right)}\equiv \int_{-\infty}^\infty d k (k)^{-2} \frac{\partial^2}{\partial \bU^i \partial \bU^j} Z.
\end{align}

We see that $Z[\bU]$ is a generating function with a non-vanishing current. In terms of Feynman diagrams this means that in addition to usual diagrams, there are now diagrams with external-current insertions.

\begin{figure}[h!]
\begingroup
\allowdisplaybreaks
\begin{alignat*}{3}
	&\begin{gathered}
		\includegraphics[width=0.2\textwidth]{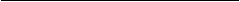}
	\end{gathered}
	&&=\omega^{-1}_{ij}
	\\
	&\begin{gathered}
		\includegraphics[width=0.2\textwidth]{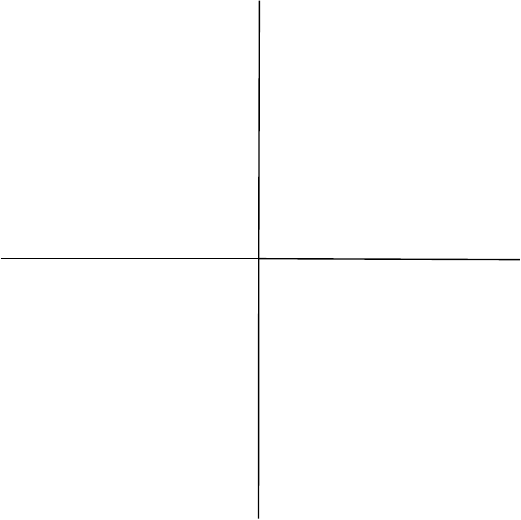}
	\end{gathered}
	&&=-\lambda_4^{ijkl}
	\\
	&\begin{gathered}
		\includegraphics[width=0.2\textwidth]{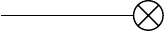}
	\end{gathered}
	&&=\delta^{ij} \kappa^{-1} \bU^{j} 
\end{alignat*}
\endgroup
	\caption{Feynman rules for the statistical prefactor.\label{fig:FeynStat}}
\end{figure}

To see the form of these external-current insertions, consider the potential 
\begin{align}
	V(x+x_B)=V_B+\frac{1}{2} x_i \omega^{ij}x_j+\frac{1}{4!} \lambda_4^{ijkl}x_i x_j x_k x_l.
\end{align}
The Feynman rules for this potential are given in Figure \ref{fig:FeynStat}.

In addition, if there are $2n$ external-current insertions, the diagram picks up a factor $(-1)^n\kappa^n(2n-1)!!$. This last rule comes from integrating over $k$. Also, diagrams with an odd-number of current insertions vanish since $\int_{-\infty}^\infty dk k^{2n+1}e^{-k^2 A}=0$.

\begin{figure}
\begingroup
\allowdisplaybreaks
\begin{alignat*}{3}
	&\begin{gathered}
	\includegraphics[width=0.2\textwidth]{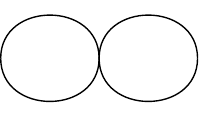}
	\end{gathered}
\quad
	&& \begin{gathered}
		\includegraphics[width=0.2\textwidth]{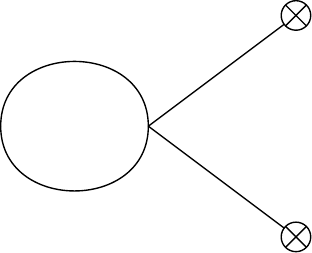}
	\end{gathered}
\quad
	&&\begin{gathered}
		\includegraphics[width=0.3\textwidth]{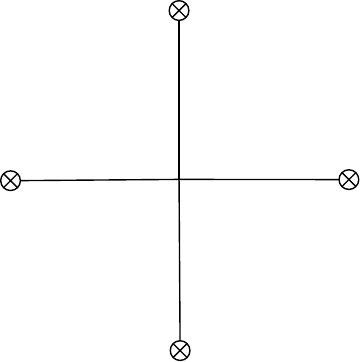}
	\end{gathered}
\\
\end{alignat*}
\endgroup
	\caption{Diagrams contributing to the statistical prefactor at NLO.\label{fig:NLODiaStat}}
\end{figure}

At next-to-leading order (NLO) the diagrams are given in Figure \ref{fig:NLODiaStat}.

Using the Feynman rules in Figure \ref{fig:FeynStat}, the first diagram in Figure \ref{fig:NLODiaStat} gives $-\frac{1}{8} \omega^{-1}_{ij}\omega^{-1}_{kl}\lambda_4^{ijkl}$; the second $\frac{1}{4} \kappa^{-1} \omega^{-1}_{ij} \bU^{k} \bU^{l} \lambda_4^{ijkl}$; and the third $-\frac{3}{4!} \kappa^{-2}\bU^i \bU^j \bU^{k} \bU^{l} \lambda_4^{ijkl}.$ 

Note that the propagator can be written
\begin{align}
	\omega^{-1}_{ij}=\sum_a \frac{\bV^i_a \bV^j_a}{\lambda_a},
\end{align}  
where $\bV_a^i$ are eigenvectors of $\omega^{ij}$ with eigenvalue $\lambda_a$.
After using this form of $\omega^{-1}_{ij}$,  the last two diagrams remove $\kappa$ from the first.

In summary, the statistical prefactor is
\begin{align}\label{eq:FinalStat}
	A_\text{stat}=e^{-\bar{S}_\text{eff}},
\end{align}
where $\bar{S}_\text{eff}$ denotes all vacuum diagrams omitting zero and negative eigenmodes. 

\subsection{Corrections to the dynamical prefactor}\label{sec:DynamicalPrefac}
Consider now the dynamical prefactor. Again, we assume a potential
\begin{align}
	V(x+x_B)=V_B+\frac{1}{2}x_i \omega^{ij}x_j+\frac{1}{3!}\lambda^{ijk}_3x_i x_j x_k+\frac{1}{4!} \lambda_4^{ijkl}x_i x_j x_k x_l.
\end{align}

Including higher-order terms in $V$ does two things. First, the Boltzmann factor changes. This is encoded in the statistical prefactor. Second, the rate of flow across the saddle-point changes. This flow is controlled by deviations from thermal equilibrium, which is encoded in the dynamical prefactor. 

Because Equation \ref{eq:ZetaEquation} describes the deviation from equilibrium, we need to find $\zeta$ to higher orders. Explicitly, we need to solve $	T\partial_i^2\zeta-\partial_i V \partial^i \zeta=0$ with the inclusion of $\lambda_3$ and $\lambda_4$ terms. The leading-order solution, denoted henceforth as $\zeta_0$, is given in Equation \ref{eq:Zeta0}.

The boundary conditions force $\zeta(\vec{x})$ to vanish as $u\rightarrow \infty$, which is enforced directly by the ansatz
\begin{align}\label{eq:ZetaGeneric}
	\zeta(\vec{x})=\zeta_0(u)+ \zeta'_0(u) \zeta_1(\vec{x})+\ldots 
\end{align}
It is assumed that $\zeta_1 =\mathcal{O}\left(\lambda_3,\lambda_4\right)$.

The equation for $\zeta_1$ can be found by combing Equations \ref{eq:ZetaGeneric} and \ref{eq:ZetaEquation}:
\begin{align}\label{eq:Zeta1}
	T	\partial_i^2 \zeta_1+\partial_i \zeta_1 \left[2\kappa u \bU^i-\omega^{ij}x_j\right]+\kappa \zeta_1=\bracket{\bU x^2}+\bracket{\bU x^3}.
\end{align}
We use the notation $\bracket{x^3}\equiv \frac{1}{2!} \lambda_3^{ijk}x_i x_j x_k$ and $\bracket{x^4}\equiv \frac{1}{3!} \lambda_4^{ijkl}x_i x_j x_k x_l$. 

All homogenous solutions of Equation \ref{eq:Zeta1} grow exponentially with $u$,  and so do not satisfy the boundary conditions. This leaves the particular solution. 

To find the particular solution, note that if we make a polynomial ansatz [in $x^i$], neither of the terms on the left-hand side of Equation \ref{eq:Zeta1} increase the polynomial's degree. Whence the ansatz go through the equations. It is also useful to work in the eigenbasis of $\omega: \omega^{ij}\bV^j_a=\lambda_a \bV^i_a$. That is, we expand $x^i=u\bU^i+\sum_a v_a \bV_a^i$ where $v_a=\bV_a^i x_i$.  This basis is useful because $\bV_a^i$ is orthogonal to $\bU^i$ and since $\omega^{ij}x_j$ reduces to a sum of eigenvalues. 

The idea is to make a polynomial ansatz in $u$ and $v_a$, where $a$ denote all eigenvalues except for the negative one. 

For simplicity, consider first the two-dimensional case $(x^i)=(x^1,x^2)$ and ignore the $\bracket{x^4}$ term for the moment. That is, we assume the potential 	
\begin{align}
V(x+x_B)=V_B+\frac{1}{2}x_i \omega^{ij}x_j	+\frac{1}{3!}\lambda^{ijk}_3x_i x_j x_k.
\end{align}
There are two eigenvectors: $\omega^{ij}\bU^j=\kappa \bU^i$ and $\omega^{ij}\bV^j=\lambda_v \bV^i$. These eigenvectors define $u\equiv \bU^i x_i$ and $v=\bV^i x_i$.

 Expanding $x^i$ in $u$ and $v$ gives
\begin{align}
	\bracket{\bU x^2}=u^2\bracket{\bU^3}+ 2 u v \bracket{\bU^2\bV }+v^2 \bracket{\bU \bV^2 }.
\end{align}

Since $\bracket{\bU x^2}$ is a degree $2$ polynomial, the relevant ansatz is $$\zeta_1=A+B_1 u+ B_2 v +C_1 u^2 +C_2 v^2 +C_3 u v.$$

After using this ansatz in Equation \ref{eq:Zeta1}, and collecting terms, one finds
\begin{alignat}{2}
	&	B_1=B_2=0, \quad &&A=-2 T( C_1+C_2)/\kappa,\nonumber
\\& C_1=\left(3\kappa \right)^{-1}\bracket{\bU^3}, \quad && C_2=\left(\kappa-2\lambda_v \right)^{-1}\bracket{\bU \bV^2},
\\& C_3=2 \left(2\kappa-\lambda_v \right)^{-1}\bracket{\bU^2 \bV}.\nonumber
\end{alignat}

Note that most of the terms in $\zeta_1$ are irrelevant. This is because the rate is given by $\int_{u=0} d^n x \bU^i J^i \sim\int  \bU^i \partial_i \zeta_1\vert_{u=0}$.
For the above example only the $C_3$ term contributes since if the derivative ($\partial^i$) hits anything else, that term is proportional to $u$ or $\bU^i \bV^i$. Both which vanish. This holds in general: we only have to care about terms containing \textit{one} factor of $u$.

The n-dimensional case is similar.
Including the quartic term in Equation \ref{eq:Zeta1}, the ansatz is a polynomial of degree $3$:
\begin{align}\label{eq:Ansatzxi}
	\zeta_1=A+ B_1 u +\sum_a B_a v_a+\ldots+ u \sum C_a v_a+\ldots+ u \sum_{a,b} D_{ab} v_a v_b+\ldots
\end{align}

Following the discussion above, only the $B_1$, $C_a$ , and $D_{ab}$ terms contribute to the rate. One finds
\begin{align}\label{eq:Ansatzxi2}
	&C_a=2 \left(2\kappa-\lambda_a \right)^{-1}\bracket{\bU^2 \bV_a},\nonumber
	\\& D_{ab}=3 \left( 2\kappa -\lambda_a -\lambda_b\right)^{-1}\bracket{\bU^2\bV_a \bV_b},
	\\& B_1=-6 T(2\kappa^{-1})\left((4\kappa)^{-1} \bracket{\bU^4}+\sum_a (2\kappa-2\lambda_a)^{-1}\bracket{\bU^2 \bV^2_a}\right).\nonumber
\end{align}
The complete solution for $\zeta_1$ is given in Appendix \ref{sec:Dim6}, and the result for potentials with $x^5$ and $x^6$ terms is also given in Appendix \ref{sec:Dim6}.

Note from Equation \ref{eq:Ansatzxi2} that $\zeta_1$ can be written in terms of Green's functions of the form
\vspace*{-0.5cm}\begin{align}\label{eq:GreenDefinition}
	\left(\omega_{ij}-c \delta_{ij}\right)G^{jk}_c=\delta_{ik} \implies G^{ij}_c=\sum_a \frac{\bV^i_a \bV^j_a}{\lambda_a-c}.
\end{align}
As before $a$ excludes the negative eigenvalue.

Before showing how $\zeta_1$ affects the rate, note that $\zeta_1$ can considered as a set of operator insertions. To see this, pick one term from Equation \ref{eq:Ansatzxi}, for example $ D_{ab} u v_a v_b$. When calculating the probability current, the factor of $u$ disappears, and the term is proportional to $v_a v_b =\bV_a^i \bV_b^j x_i x_j$. Adding this term to Equation \ref{eq:LeadingOrderRate},  one could equally well have inserted an operator $\mathcal{O}\propto D_{ab}\bV_a^i \bV_b^j x_i x_j$. This operator can also receive loop corrections like in normal field theory.

Additional corrections to $\zeta$ can be found by using $\zeta=\zeta_0+\zeta_1 \zeta'_0+ \zeta_2 \zeta'_0+\ldots$. Where it is assumed that $\zeta_n =\mathcal{O}\left(\lambda^n\right)$. In general, $\zeta_n$ is a polynomial of degree $4 n-1$, and is completely determined by $\zeta_{n-1}$. Namely, for $n>0$
\begin{align}
	T	\partial_i^2 \zeta_n+\partial_i \zeta_n \left[2\kappa u \bU^i-\omega^{ij}x_j\right]+\kappa \zeta_n=&\left(\kappa \frac{u}{T}\zeta_{n-1}+\partial_u  \zeta_{n-1} \right)\left(\bracket{\bU x^2}+\bracket{\bU x^3}\right)\nonumber
	\\& + \sum_a \frac{\partial}{\partial v_a}\zeta_{n-1} \left(\bracket{\bV^a x^2}+\bracket{\bV^a x^3}\right)
\end{align}

\begin{figure}[h!]
	\centering
	\includegraphics[scale=0.75]{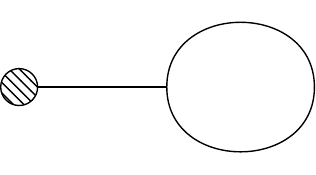}
	\caption{Tadpole contribution to the dynamical prefactor.\label{fig:Tadpole}}
\end{figure}

Naively one would think that $\zeta_2$ is sub-leading compared to $\zeta_1$. However, from Equation \ref{eq:Ansatzxi} we see that the $c_a$  term does not contribute directly to the rate. Instead this term contributes via the one-loop tadpole shown in Figure \ref{fig:Tadpole}. 
	
\vspace*{-0.5cm}
\vspace*{-0.5cm}	
	
This tadpole insertion scales as $\lambda_3^2$, and similar terms exist in $\zeta_2$. As such, all $\lambda_3^2$ terms in $\zeta_2$ are relevant. 

 We here only note that these terms are of the form
\begin{align}
	\zeta_2= u B' +u D'_{ab} v_{a}v_b+u F_{abcd}v_{a}v_b v_c v_d+\ldots	
\end{align}
The explicit expressions are given in Equation \ref{eq:NLOCubic}

In summary, the dynamical prefactor consists of operator insertions, which multiply the full statistical prefactor.

\subsection{Comparison with known results}
Sections \ref{sec:StatPrefactor} and \ref{sec:DynamicalPrefac} derived the nucleation-rate for systems with $n$ degrees of freedom. There are however pre-existing results for $n=1$. Take a potential
\begin{align}
V=-\frac{1}{2}\kappa x^2+\frac{1}{3!}\lambda_3 x^3+ \frac{1}{4!} \lambda_4 x^4.
\end{align}
The NLO rate is\cite{Larson1.436510,RevModPhys.62.251}:
\begin{align}\label{eq:1DResult}
	\Gamma=\frac{\sqrt{\abs{\kappa}}}{2\pi \eta}\left[1-T\left(\frac{\lambda_4}{8 \kappa^2}-\frac{5 \lambda_3^2}{24\kappa^3}\right)\right]e^{-V_B/T},
\end{align}
where as before the normalization from the metastable state is left implicit. This formula includes all corrections to the dynamical and statistical prefactors.

It is encouraging that the statistical prefactor in Equation \ref{eq:FinalStat} together with the dynamical prefactor in Equation \ref{eq:Ansatzxi2} (and $\xi_2$ from Equation \ref{eq:NLOCubic}) reproduce Equation \ref{eq:1DResult}.

\section{Generic damping coefficients}\label{sec:GeneralDamping}
Consider now the generic-damping case. There are a few new features as compared to the overdamped system. First, the velocity, or rather the conjugate momenta, contribute to the Boltzmann factor. Second, the damping is not solely a multiplicative factor as in Equation \ref{eq:LeadingOrderPrefactor}. Third, the negative-eigenvalue direction contributes to the Boltzmann integral.

As before, the starting point is the Fokker-Planck equation\cite{Fokker,Langer:1969bc,Berera:2019uyp}:
\begin{align}
	\frac{d}{d t}P=-\partial^i_x J_x^i-\partial^i_\pi J_\pi^i=\left[-\pi_i \partial_x^i+\partial_\pi^i\left[\eta \pi ^i+\partial_x^i V\right]+\eta T \partial_\pi^i\right]P=0,
\end{align}
here $\pi^i$ are conjugate momenta associated with $x^i$, and $\eta$ is the damping.

To derive the rate, we start by assuming
\vspace*{-0.3cm}\begin{align}
	V(x+x_B)\approx V_B+\frac{1}{2}x_i \omega^{ij} x_j.
\end{align} 
Next, we make the ansatz $P=\zeta(\vec{x},\vec{v})P_0$ where $P_0=e^{-E/T}$ and $E=\frac{1}{2} \pi_i^2+V(x)$. Hence
\begin{align}\label{eq:ZetaEqGeneral}
	\left[	\pi^i \partial_x^i+\left[\eta \pi ^i-\omega^{ij}x_j\right]\partial_\pi^i-\eta T \partial_\pi^2\right] \zeta=0.
\end{align}
Using the ansatz $\zeta(\vec{x},\vec{v})=\zeta(u)$, where $u=\bU_\pi^i \pi^i+\bU_x^i x^i$, the equation becomes
\begin{align}\label{eq:ZetaEqGeneralAnsatz}
	\pi^i \bU_x^i \zeta'(u)+\bU_\pi^i \left[\eta \pi^i-\omega^{ij} x_j\right] \zeta'(u)-\eta T \bU_\pi^2 \zeta''(u)=0.
\end{align}
For consistency
\vspace*{-0.3cm}\begin{align}
	\pi^i \bU_x^i +\bU_\pi^i \left[\eta \pi^i-\omega^{ij} x_j\right]=-\lambda u=-\lambda \left[ \bU_\pi^i \pi^i+\bU_x^i x^i\right].
\end{align}
This implies that $\bU_\pi^i$ is an eigenvector of $\omega^{ij}$: $\bU_\pi^i \omega^{ij} x_j=\kappa \bU_\pi^i x_i$. 

Thus
\vspace*{-0.3cm}\begin{align}\label{eq:GeneralLambda}
	-\lambda \bU_\pi^i=\left(\bU_x^i+\eta \bU_\pi^i\right) \hspace{0.2cm}  \& \hspace{0.2cm} -\kappa \bU_\pi^i=-\lambda \bU_x^i,
\end{align}
so $\bU_x^i$ is also an eigenvector of $\omega^{ij}$.

Putting everything together one finds\cite{Langer:1969bc,RevModPhys.62.251}
\begin{align}
	\lambda=\frac{1}{2} \left(\pm\sqrt{\eta ^2-4 \kappa }-\eta \right).
\end{align}

It will be shown below that $\lambda$ needs to be positive, so perforce $\kappa<0$ and we must choose $\lambda=\frac{1}{2} \left(\sqrt{\eta ^2-4 \kappa }-\eta \right)$. Next, using Equations \ref{eq:ZetaEqGeneralAnsatz} and \ref{eq:GeneralLambda} gives~\cite{Langer:1969bc,RevModPhys.62.251}
\begin{align}\label{eq:Sol0General}
	&\gamma \zeta''(u)+u\zeta'(u)=0,\quad \gamma=\eta T \frac{\lambda}{\kappa^2}>0.
\end{align}

The solution satisfying the boundary conditions is
\begin{align}\label{eq:GeneralZeta}
	\zeta(u)=\frac{1}{\sqrt{2 \pi \gamma}}\int_u^\infty e^{-\frac{u'^2}{2\gamma}} du'
\end{align}

Given $\zeta(u)$, it is straightforward to calculate the rate.
Explicitly\cite{Langer:1969bc,RevModPhys.62.251}
\begin{align}\label{eq:BoltzmanGeneral}
	\Gamma=\frac{\bU_\pi^2 \eta}{\sqrt{2 \pi \gamma}} \int_{u=0} d^n x d^n\pi e^{-E/T},
\end{align}
where $E=\frac{1}{2} \pi_i^2+V(x)$.

Doing the Gaussian integrals one finds up to a normalization 
\begin{align}
	\Gamma=\frac{\lambda}{2\pi} \left\vert \text{det}' \omega \right\vert^{-1/2}e^{-V_B /T}.
\end{align}

As before there are two types of contributions from higher orders. First, corrections to $\zeta$, and second, higher-order corrections to $E$ in the integral $ \int_{u=0} d^n x d^nv e^{-E/T}$.

Before calculating these corrections, note that the rate is still real. To see this, denote the negative-eigenvalue direction by $x_u\equiv \bU_x^i x_i$. The problematic term in the energy is\vspace*{-0.3cm}
\begin{align}
	E=  -\frac{1}{2}\abs{\kappa}x_u^2+\ldots
\end{align} 
If the rate is calculated at $u=0$, we have $\pi_u=\bU_\pi^i \pi^i=\frac{\abs{\kappa}}{\lambda} x_u$. Hence
\begin{align}\label{eq:GeneralModEig}
	-\abs{\kappa}\rightarrow \frac{\kappa^2}{\lambda^2}-\abs{\kappa} =\frac{1}{2} \eta  \left(\sqrt{\eta ^2+4 \abs{\kappa} ^2}+\eta \right)>0.
\end{align}

We see that the Gaussian integral in Equation \ref{eq:BoltzmanGeneral} converges, and that the rate is real.

\subsection{Physical interpretation and applicability}
The rate is calculated at $u=0$. Let us now see what this means. For clarity we use $x_u\equiv \bU_x^i x_i$ and $\pi_u\equiv \bU_\pi^i \pi^i$. That is, $x_u$ and $\pi_u$ are the position and conjugate momentum in the negative-eigenvalue direction. Denote the energy in this direction by $E_u\equiv \frac{1}{2}\pi_u^2+\frac{1}{2}\kappa x_u^2$. Consider now the motion of a particle starting at the saddle point. To do so, recall the static Fokker-Planck equation:
\begin{align}\label{eq:FPFoward}
	\left[\pi_i \partial_x^i-\partial_\pi^i\left[\eta \pi ^i+\partial_x^i V\right]-\eta T \partial_\pi^i\right]P=0.
\end{align}
In addition, if $P=\zeta(\vec{x},\vec{\pi})e^{-E/T}$, the equation for $\zeta(\vec{x},\vec{\pi})$ is
\begin{align}\label{eq:FPBackward}
	\left[	\pi^i \partial_x^i+\left[\eta \pi ^i-\partial^i_x V\right]\partial_\pi^i-\eta T \partial_\pi^2\right] \zeta=0.
\end{align}
Namely, deviations from thermal equilibrium follow Equation \ref{eq:FPBackward}, which is known as the adjoint Fokker-Planck equation.

Now, the Fokker-Planck equation corresponds to a particle obeying
\begin{align}\label{eq:Langevin}
	\dot{\pi}^i=-\eta \pi^i-\partial_x^i V+f(t),
\end{align}
the last term represents thermal noise and vanishes on average.

Likewise, the backward Fokker-Planck equation follows from
\begin{align}\label{eq:LangevinBackward}
	\dot{\pi}^i=\eta \pi^i-\partial_x^i V+f(t),
\end{align}
Imagine now releasing a particle close to $x_u=\pi_u=0$. If the particle evolves according to Equation \ref{eq:Langevin}, an unstable solution is\cite{Langer:1969bc}
\begin{align}
	x^i=x_0 e^{\lambda t} \quad \pi^i=x_0 \lambda e^{\lambda t} .
\end{align}
On this solution the energy is\vspace*{-0.3cm}
\begin{align}\label{eq:EnergyFP}
	E_u=-\frac{1}{2}\eta \lambda x_0^2 e^{2 t \lambda}.
\end{align}
As expected $\frac{d}{dt}E=-\eta v^2<0$.

Next, consider instead a particle following Equation \ref{eq:LangevinBackward}. The unstable solution is now
\begin{align}
	x^i=x_0 e^{\frac{\ak}{\lambda} t} \quad \pi^i=x_0 \frac{\ak}{\lambda}e^{\frac{\ak}{\lambda} t},
\end{align}
with associated energy\vspace*{-0.3cm}
\begin{align}\label{eq:EnergyFPBackward}
	E_u=\frac{1}{2}\eta \frac{\ak}{\lambda} x_0^2 e^{2\frac{\ak}{\lambda} t}.
\end{align}
This is precisely the solution implied by $u=0$: $\pi_u=\frac{\ak}{\lambda} x_u$. Note that the energy in Equation \ref{eq:EnergyFPBackward} suppresses fluctuations away from the saddle point via the Boltzmann factor. This is why the Boltzmann-integral is localized at $x_u=0$ for $\eta \rightarrow \infty$.

Equation \ref{eq:EnergyFPBackward} also highlights a problem for small damping. Namely, for $\eta=0$ the energy in Equation \ref{eq:EnergyFPBackward} identically vanishes. Effectively the negative-eigenmode turns into a zero-mode. While the rate is finite in the $\eta \rightarrow 0$ limit to leading order, this does not generalize to higher orders.

There are other indications that the formalism breaks down for small damping\cite{RevModPhys.62.251}, and one should be cautious when using existing formulas\cite{Linde:1981zj,Affleck:1980ac} for an underdamped system.

\subsection{The statistical prefactor}

Consider now corrections to the statistical prefactor. As in Section \ref{sec:StatPrefactor} we start with the integral
\begin{align}
	Z[ \bU_x,\bU_\pi]=	\int d^n x d^n \pi e^{i k u-1/T\frac{1}{2} \left(\pi^i\pi^i+x_i \omega^{ij}x_j\right)},
\end{align}
where $u=\bU_x^i x_i +\bU_\pi^i \pi^i$.

One finds
\begin{align}\label{eq:GeneratingFunctional}
	Z[\bU_x,\bU_\pi]=\left[\text{det}' \abs{\omega}/(2\pi)\right]^{-1/2}(2\pi)^{-n/2}\exp\left[\frac{1}{2}k^2 T\left(\bU_\pi^2 +\bU_x^i \omega^{-1}_{ij} \bU_x^j\right)\right].
\end{align}

We see that $	Z[\bU_x,\bU_\pi]$ is a generating function, and the rule is
\begin{align}
	x^i \rightarrow k^{-1}\frac{\partial}{\partial \bU_x^i}Z[\bU_x,\bU_\pi],\quad \pi^i \rightarrow k^{-1} \frac{\partial}{\partial \bU_\pi^i}Z[\bU_x,\bU_\pi].
\end{align}

The integral converges because
\begin{align}
-T\left(\bU_\pi^2 +\bU_x^i \omega^{-1}_{ij} \bU_x^j\right)=\frac{T \eta \lambda}{\kappa^2} =\gamma>0,
\end{align}
with $\gamma$ given in Equation \ref{eq:Sol0General}.

\begin{figure}
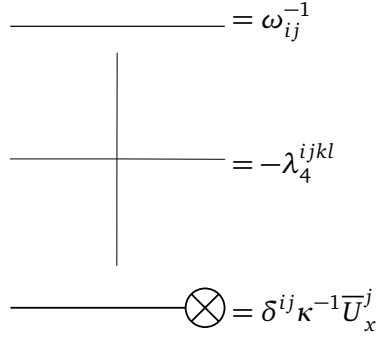

\begingroup
\allowdisplaybreaks
\begin{alignat*}{3}
	&\begin{gathered}
		\includegraphics[width=0.2\textwidth]{propagator}
	\end{gathered}
	&&=\omega^{-1}_{ij}
	\\
	&\begin{gathered}
		\includegraphics[width=0.2\textwidth]{quartic}
	\end{gathered}
	&&=-\lambda_4^{ijkl}
	\\
	&\begin{gathered}
		\includegraphics[width=0.2\textwidth]{ExternalCurrent}
	\end{gathered}
	&&=\delta^{ij} \kappa^{-1} \bU_x^{j} 
\end{alignat*}
\endgroup
	\caption{Feynman rules for general damping.\label{fig:FeynOverDamp}}
\end{figure}

As before, consider the potential 
\begin{align}
	V(x+x_B)=V_B+\frac{1}{2} x_i \omega^{ij}x_j+\frac{1}{4!} \lambda_4^{ijkl}x_i x_j x_k x_l.
\end{align}
The Feynman rules for this potential are given in Figure \ref{fig:FeynOverDamp}.

In addition to these rules, if there are $2n$ external-current insertions, the diagram picks up a factor $(-1)^n\gamma^{-n}(2n-1)!!$; diagrams with an odd number of current insertions vanish. 

Similarly, for integrals over the conjugate momenta, the external current gives a factor $\delta^{ij}\bU_\pi^j=\frac{\lambda}{\kappa}\delta^{ij}\bU_x^j$, and the $2n$ rule is the same.

The derivation of the dynamical prefactor is given in Appendix \ref{app:GeneralDynamical}.

\section{Field theory}\label{sec:FieldTheory}
As yet we have not considered field theory, however, the field-theory limit is straightforward \cite{Berera:2019uyp}. From Section \ref{sec:NucleationRate} we known that these fields live in three dimensions. Furthermore, as discussed in Section \ref{sec:HighT}, high-temperature effects can be captured by using effective parameters in this theory~\cite{Gould:2021ccf,Bodeker:1999ey}. Yet for the purposes of this paper we do not consider an explicit effective theory~\cite{Farakos:1994kx,Kajantie:1995dw}, and three-dimensional parameters are taken as free. We will however absorb all temperature-dependence by rescaling couplings and fields. In addition, we omit details associated with the assumed, effective, Langevin description. Uncertainties associated with this description are important\cite{Blaizot:1993be,Blaizot:1993zk,Berera:2007qm,Gautier:2012vh,Moore:2000jw,Bodeker:1999ey,Arnold:1998cy}, but lie beyond the scope of this paper.

Consider a real-scalar model with three-dimensional action
\begin{align}
	S_\text{LO}[\phi]=\int d^3 x\left[\frac{1}{2}(\vec{\nabla} \phi)^2+V(\phi) \right].
\end{align}

The barrier position $\vec{x}_B$ is replaced with the bounce $\phi_b(\abs{\vec{x}})$~\cite{Coleman:1977py,Linde:1981zj}; the action evaluated on the bounce is analogous to the barrier height: 
\begin{align}
	V_B\rightarrow S_\text{LO}[\phi_b].
\end{align}
In Section \ref{sec:NucleationRate} we assumed that
\begin{align}
	V(x+x_B)=V_B +\frac{1}{2}x_i\omega^{ij}x_j+\frac{1}{3!} \lambda_3^{ijk} x_i x_j x_k+\frac{1}{4!} \lambda_4^{ijkl} x_i x_j x_k x_l.
\end{align}

In field theory this potential is replaced with (to fourth order)
\begin{align}
	\SLO[\phi_b+\phi]&=\SLO[\phi_b]+\frac{1}{2}\int d^3y d^3 w \phi(y)\delta_y \delta_w \SLO[\phi_b] \phi(w)
	\\&+ \frac{1}{3!}\int d^3y d^3 w d^3z \delta_y \delta_w \delta_z \SLO[\phi_b]\phi(y) \phi(w) \phi(z) 
	\\&+\frac{1}{4!}\int d^3y d^3 w d^3zd^3v \delta_y \delta_w \delta_z \delta_v \SLO[\phi_b] \phi(y) \phi(w) \phi(z) \phi(v),
\end{align}
where $\delta_y$ is shorthand for $\frac{\delta}{\delta \phi(y)}$.

We see that the dictionary is
\begin{align}
	&	x_i\omega^{ij}x_j \rightarrow \int d^3 x \phi(x) \left[-\partial^2+V''[\phi_b]\right]\phi(x)
	\\& \lambda_3^{ijk} x_i x_j x_k\rightarrow \int d^3x V'''[\phi_b] \phi(x)^3,
	\\& \lambda_4^{ijkl} x_i x_j x_k x_l\rightarrow \int d^3x V''''[\phi_b] \phi(x)^4,
\end{align}
where $\partial^2\equiv \vec{\nabla}^2$.

Furthermore, the solution to the Fokker-Planck equation is~\cite{Berera:2019uyp} 
 \begin{align}
	P=\zeta(u)e^{-\SLO[\phi+\phi_b]},
\end{align}
where for an overdamped system $\zeta(u)$ is given by
\begin{align}
	&\zeta(u)=\frac{1}{\sqrt{2\pi \abs{\alpha}}}\int_{u}^\infty du' e^{-u'^2/(2\abs{\alpha})},\quad \alpha=\frac{1}{\kappa},
\end{align}
 and $u= \int d^3x \bU(x) \phi(x)$ is defined by
\begin{align}
	\left[-\partial^2+V''[\phi_b] \right]\bU(x)=\kappa \bU(x),\quad \kappa<0.
\end{align}

\subsection{Feynman rules for the statistical prefactor}\label{sec:FeynmanRules}

Consider the potential
\begin{align}\label{eq:RealScalar}
	 V(\phi)=\frac{1}{2}m^2\phi^2-\frac{1}{3!}g \phi^3+\frac{1}{4!}\lambda \phi^4.
\end{align}
  Expanding around the bounce gives\footnote{Linear terms are omitted since $\delta S_\text{LO}[\phi_b]=0$.}
\begin{align}
	V(\phi_b+\phi)=&V(\phi_b)+\frac{1}{2!}\phi^2\left[m^2-g\phi_b+\frac{1}{2}\lambda \phi_b^2 \right]\nonumber
	\\& + \frac{1}{3!}\left[\lambda \phi_b-g\right]\phi^3+ \frac{1}{4!}\lambda \phi^4.
\end{align}

\begin{figure}
\begingroup
\allowdisplaybreaks
\begin{alignat*}{3}
	&\begin{gathered}
		\includegraphics[width=0.2\textwidth]{propagator}
	\end{gathered}
	&&=\Delta(x-y)
	\quad \begin{gathered}
		\includegraphics[width=0.2\textwidth]{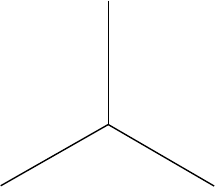}
	\end{gathered}
	&&=-(\lambda \phi_b(x)-g)
	\\& \begin{gathered}
		\includegraphics[width=0.2\textwidth]{quartic}
	\end{gathered}
	&&=-\lambda 
	\quad \quad \begin{gathered}
		\includegraphics[width=0.2\textwidth]{ExternalCurrent}
	\end{gathered}
	&&= \kappa^{-1} \bU(x),
\end{alignat*}
\endgroup
	\caption{Feynman rules for a real-scalar theory.\label{fig:FeynRealScalar}}
\end{figure}

Using the result in Section \ref{sec:StatPrefactor}, we obtain the Feynman rules given in Figure \ref{fig:FeynRealScalar}.

 The propagator is defined by  $\left[-\partial^2+V''[\phi_b(x)]\right]\Delta(x-y)=\delta^{3}(x-y)$; all zero eigenvalues are projected out as explained in Section \ref{sec:ZeroModes}. Also, this propagator depends on the bounce, and needs to be found numerically. See Section \ref{sec:PracticalGreen} for the details.

The Feynman rules work as usual, with the addition that if there are $2n$ external-current insertions, the diagram picks up an additional factor $(-1)^n\kappa^n(2n-1)!!$, and diagrams with an odd number of current insertions vanish.

\begin{figure}[h!]
	\includegraphics[scale=0.7]{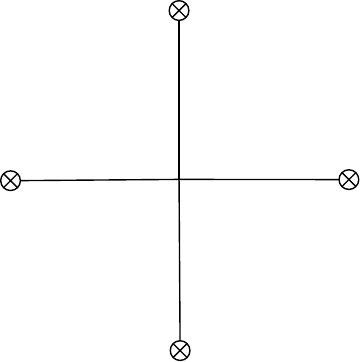}
		\caption{Diagram with 4 external-current insertions.\label{fig:RealScal4ptExt}}
\end{figure}

As an example, consider the diagram in Figure \ref{fig:RealScal4ptExt}.

Using the Feynman rules given in Figure \ref{fig:FeynRealScalar} we find
\begin{align}
	(-\lambda)/4! \kappa^{-4} (-1)^2\kappa^2 (4-1)!! \int d^3x \bU^4(x) =-3/4!\lambda \kappa^{-2}\int d^3x \bU^4(x).
\end{align}

\begin{figure}[h]
\begingroup
\allowdisplaybreaks
\begin{alignat*}{3}
	&\begin{gathered}
		\includegraphics[width=0.2\textwidth]{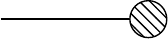}
	\end{gathered}
	&&= \sum_{a}\bV_a(x) \left(\lambda_a-2\kappa \right)^{-1}\int d^3w \bU^2(w) \bV_a(w) \left(\lambda \phi_b(w)-g\right)\\
	&\begin{gathered}
		\includegraphics[width=0.2\textwidth]{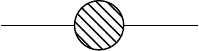}
	\end{gathered}
	&&=  \frac{1}{2}\lambda \sum_{ab}\bV_a(x)\bV_b(y) \left(\lambda_a+\lambda_b-2\kappa \right)^{-1}\int d^3w \bU^2(w) \bV_a(w)\bV_b(w)
\end{alignat*}
\begin{alignat*}{2}
&\begin{gathered}
		\includegraphics[width=0.04\textwidth]{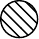}
	\end{gathered}
	&&= \lambda \frac{1}{4}\kappa^{-1}\int d^3x \left[(2\kappa)^{-1}\bU^4(x)-\Delta_\kappa(x,x)\bU^2(x) \right],
\end{alignat*}
\endgroup
	\caption{Dynamical prefactor Feynman rules.\label{fig:RealScalFeynDyn}}
\end{figure}

The dynamical prefactor is calculated via operator insertions as explained in Section \ref{sec:DynamicalPrefac}. These operators are given in Figure \ref{fig:RealScalFeynDyn}.

The first vertex acts as a tadpole, the second as a $2$-point insertion, and the third is a number. Note that the second vertex can be reduced to a sum of propagators once it contracts with other lines. For example, because $\bV_a(x)$ are eigenfunctions of the propagator, we have
\begin{align}
	\int d^3 y \Delta(x,y)\bV_a(y)=\bV_a (x) \lambda_a^{-1}.
\end{align}

\begin{figure}[h!]
	\includegraphics[scale=0.6, trim = {0 0.6cm 0 0.9cm}]{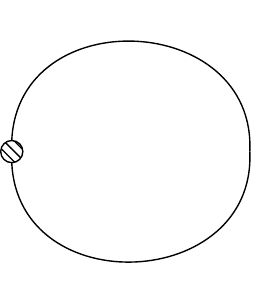}
	\caption{Diagram contributing to the dynamical prefactor.\label{fig:RealScal2ptBubble}}
\end{figure}

Then the leading correction involving the $2$-point insertion is given in Figure \ref{fig:RealScal2ptBubble}.

After using the Feynman rules we find
\begin{align}\label{eq:QuarticDynamical}
	&	\frac{\lambda}{4}  \sum_a \lambda_a^{-1}\frac{1}{\lambda_a-\kappa}\int d^3w \bU^2(w)\bV^2_a(w) =\nonumber
	\\&\frac{\lambda}{4\kappa}\int d^3 w\left[\Delta_\kappa(w,w)\bU^2(w)-\Delta(w,w)\bU^2(w)+\bU^4(w)\kappa^{-1}\right],
\end{align}
where the shifted propagator satisfies
\begin{align}\label{eq:ShiftedProp}
\left[-\partial^2+V''[\phi_b(x)]-c\right]\Delta_c(x-y)=\delta^{3}(x-y)
\end{align}

\subsection{Zero-modes}\label{sec:ZeroModes}
We now turn to zero-modes. In field theory there are three zero-modes associated with the bounce\cite{Coleman:1978ae,Coleman:1977py}. These zero-modes are readily identified: $\bV_\mu(x)=\left(S_B\right)^{-1/2}\partial_\mu \phi_b(x)$~\cite{Andreassen:2016cvx,Coleman:1978ae,GERVALS1976281}.

We can remove these zero-modes by using collective coordinates~\cite{Andreassen:2016cvx,GERVALS1976281} 
\begin{align}
	\phi(x)=\sum_a v_a \bV_a (x-x_0),
\end{align}
where $a$ excludes zero-modes. Essentially the zero-modes turn into a volume factor. In particular, by using collective coordinates the zero-modes \textit{never} appear in propagators. 

 Using collective coordinates also gives a Jacobean. To leading order this Jacobean is $\left(S_B\right)^{3/2}$. However, as emphasized in \cite{Andreassen:2016cvx}, there can be additional terms. To see the effect of these terms, one can extend the result of \cite{Andreassen:2016cvx} to a three-dimensional field theory. The result is
\begin{align}\label{eq:ZeroModeOperator}
	J_{ZM}=& \left(S_B\right)^{3/2}\left\vert 1-(S_B)^{-1} \int d^3x \phi(x) \partial^2 \phi_b(x)\right.
	\\& \left.+(S_B)^{-2}\frac{1}{2}\int d^3x d^3y \phi(x)\phi(y)\left[\partial^2 \phi_b(x)\partial^2 \phi_b(y)-\partial_\mu \partial_\nu \phi_b(x)\partial_\mu \partial_\nu \phi_b(y)\right] \right.\nonumber
	\\& \left. -(S_B)^{-3}A_3 \right\vert  ,\nonumber
\end{align}
where\footnote{The second term in $A_3$ can be symmetrized over $x,y$ and $z$.}
\begin{align}
	A_3=&\int d^3x d^3y d^3z \phi(x)  \phi(y) \phi(z)\left[\frac{1}{3}\partial_\mu \partial_\nu \phi_b(x)\partial_\nu \partial_\alpha \phi_b(y)\partial_\alpha \partial_\mu \phi_b(z) \right.\nonumber
	\\& \left.-\frac{1}{2}\partial^2 \phi_b(x)\partial_\mu \partial_\nu \phi_b(y)\partial_\mu \partial_\nu \phi_b(z) +\frac{1}{6}\partial^2\phi_b(x)\partial^2\phi_b(y)\partial^2\phi_b(z)
	\right].
\end{align}

The first term in $J_{ZM}$ gives the familiar prefactor $(S_B)^{3/2}$, which cancels when calculating the statistical prefactor\cite{Dunne:2005rt,Ekstedt:2021kyx}. The other terms are operator insertions that need to be taken into account order-by-order. Note that these zero-mode operators talk with the dynamical prefactor.

\subsection{Connection with the effective action}\label{sec:EffectiveAction}
It is sometimes useful to view the rate as an effective action.  Take for example the statistical prefactor
\begin{align}\label{eq:EffectiveAction}
	A_\text{stat}=\int_{u=0} \mathcal{D} \phi e^{-S[\phi+\phi_b]}=e^{-\overline{S}_\text{eff}[\phi_b]}.
\end{align}

This effective action is the sum of all vacuum diagrams in the bounce background. In getting to the right-hand side, we have assumed that $\phi_b$ is a generic background field. To obtain the rate one should fix $\phi_b$ so that $\delta S_\text{eff}[\phi_b]=0$.

\begin{figure}[h!]
	\includegraphics[scale=0.5]{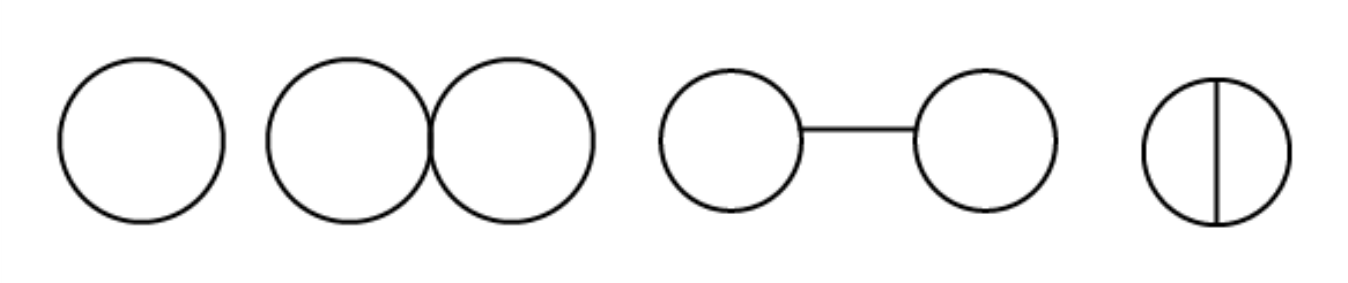}
	\caption{Diagrams contributing to the statistical prefactor.\label{fig:RealScalEffAc2Loop}}
\end{figure}

We could of course calculate the integral in Equation \ref{eq:EffectiveAction} numerically without any reference to the effective action. However, the effective action can be useful. On the practical level the effective-action approach removes tadpole diagrams. For example, omitting external-current insertions, the statistical prefactor in Equation \ref{eq:EffectiveAction} is given by the diagrams in Figure \ref{fig:RealScalEffAc2Loop}.

The third diagram is not part of the effective action. Instead, corrections to the leading-order bounce incorporate all tadpoles.
Explicitly, if the tree-level action is $S_\text{LO}$ and the one-loop correction is $S_\text{NLO}$, then\footnote{Integrations are left implicit. }
\begin{align}
	&	\delta S_\text{LO}[\phi_\text{LO}]=0 \hspace{ 3.8cm} \left( \text{The bounce solution}\right)
	\\& \phi_\text{NLO} \delta^2 S_\text{LO}[\phi_\text{LO}]+ \delta S_\text{NLO}[\phi_\text{LO}]=0  \hspace{ 0.5cm} \left( \text{One-loop correction to the bounce}\right)
\end{align}

Including $\phi_\text{NLO}$ in the effective action gives
\begin{align}
	&S_\text{eff}[\phi_b]=\SLO[\phi_b]+\SNLO[\phi_b]+\ldots
	\\&\implies \SLO[\bLO]+\left[\SNLO[\bLO]-\frac{1}{2} \bNLO^2\delta^2 \SLO[\bLO]\right]
\end{align}
Diagrammatically $\SNLO[\bLO]$ comes from the double-bubble and sunset diagrams, and $-\frac{1}{2} \bNLO^2\delta^2 \SLO[\bLO]$ is the dumbbell diagram.

The actual expression for $\bNLO$ is given by
\begin{align}
	\bNLO(x)=-\int d^3y \Delta(x,y)\frac{\delta \SNLO[\bLO]}{\delta \phi(y)}.
\end{align}
This follows from the definition $\bNLO$:
\begin{align}
	-\left[-\partial^2+V''[\bLO(x)]\right]\bNLO(x)=\frac{\delta \SNLO[\bLO]}{\delta \phi(x)}.
\end{align}
Note that both the propagator, and implicitly the effective action, are defined with zero-modes projected out. 

\subsection{Renormalization-scale invariance of the rate}\label{sec:RG}
Calculations should be independent of the renormalization scale. To show that the rate is scale-independent is however non-trivial.

Note that renormalization-scale dependence first shows up first at two loops in three dimensions. Moreover, the two-loop beta function is the complete beta function, and only mass parameters depend on the scale in three dimensions. For the real-scalar model considered in Equation \ref{eq:RealScalar}, the beta function is
\begin{align}
	\frac{d m^2}{d\log \mu_3}=\frac{1}{16 \pi^2} \frac{1}{3!} \lambda^2.
\end{align}

Before considering the rate, it is instructive to first calculate the effective potential.
 That is, we treat $\phi_b\rightarrow \phi$ as a constant background field.

\begin{figure}[h!]
	\includegraphics[scale=0.4]{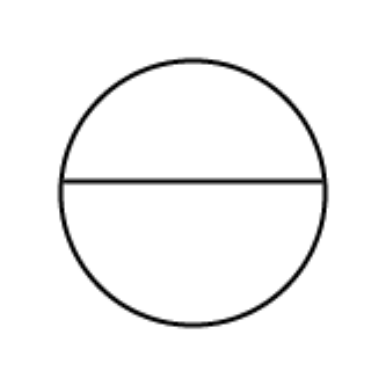}
		\caption{Diagram contributing to the two-loop effective potential.\label{fig:RealScalEffPot2Loop}}
\end{figure}

The relevant diagram is given in Figure \ref{fig:RealScalEffPot2Loop}.

Using normal Feynman rules, the contribution to the effective potential is
\begin{align}
	-\frac{1}{3!\times 2!}(\lambda \phi-g)^2 \int d^3x d^3y \Delta(x-y)^3=-\frac{1}{3!\times 2!}(\lambda \phi^2-g) \int d^3x d^3y \Delta(x)^3,
\end{align}
where the last step only works if the background field is constant. The propagator in the background field is defined by
\begin{align}
	\left[-\partial^2+M^2\right]\Delta(x-y)=\delta^{(d)}(x-y), \hspace{0.5cm} M^2\equiv V''(\phi).
\end{align}

The solution, in $d=3-2\epsilon$ dimensions, is \cite{Braaten:1995cm}
\begin{align}
	 \Delta(x)=\left(\frac{e^\gamma \mu_3^2}{4\pi}\right)^\epsilon \frac{1}{(2\pi)^{3/2-\epsilon}}\left(\frac{M}{\abs{x}}\right)^{1/2-\epsilon}K_{1/2-\epsilon}(M \abs{x}),
\end{align}
 here $K_{1/2-\epsilon}(x)$ is the modified Bessel function. Note that for $\epsilon=0$ the propagator is 
\begin{align}
	\Delta(x)=\frac{e^{- M \abs{x}}}{4\pi \abs{x}}.
\end{align}

Note that all $\epsilon$ poles, and associated scale-dependence, come from the $\abs{x}\rightarrow 0$ region. While contributions from the $\abs{x}\rightarrow \infty$ region are finite. The idea is to introduce a radial cut-off $R$, and for $\abs{x}<R$ one should use the $d=3-2\epsilon$ expression for the propagator, while for $\abs{x}\geq R$ one can directly take the $\epsilon\rightarrow 0$ limit \cite{Rajantie:1996np}. Using properties of the Bessel function one finds
\begin{align}\label{eq:EffPotInt}
	\int_{\abs{x}<R}	d^3x \Delta(x)^3=\frac{4 \epsilon  \log (\mu_3  R)+(2+4 \gamma ) \epsilon +1}{64 \pi ^2 \epsilon }.
\end{align}
Note that the $\log R$ term cancels when including the region $\abs{x}\geq R$ \cite{Rajantie:1996np}.

Consider the scale-dependent piece:
\begin{align}
	\frac{d V_\text{2-Loop}}{d\log \mu_3}=-\frac{1}{96\pi^2\times 2!}\lambda^3 \phi^2,
\end{align}
which cancels with the running of the tree-level mass:
\begin{align}
	\frac{d}{d\log\mu_3}\left(\frac{1}{2}m^2\phi^2\right)=\frac{1}{16\pi^2}\frac{1}{2\times 3!}\lambda^2 \phi^2.
\end{align}

With the effective-potential case done, most steps carry over to the nucleation rate. First, note that the scale dependence of the leading-order bounce action is\footnote{$	\frac{d}{d\log\mu_3}\phi_b$ terms cancel since the action is evaluated on-shell.}
\begin{align}
	\frac{d}{d\mu_3}\SLO[\phi_b]=\int d^3x \frac{1}{2}\phi_b^2\frac{d}{d\log\mu_3} m^2.
\end{align}

The sunset diagram gives
\begin{align}\label{eq:SunsetRate}
	-\frac{1}{3!\times 2!}\lambda^2\int d^3x d^3y \Delta(x,y)^3 \phi_b(x)\phi_b(y).
\end{align}
We omit terms linear in $\phi_b$ as they correspond to tadpoles.

The region of interest is $x\approx 0$, and in this limit the propagator satisfies\footnote{The same result follows after explicitly calculating the propagator as shown in Section \ref{sec:PracticalGreen}.}
\begin{align}\label{eq:DerivExp}
	&	\left[-\partial^2+V''(\phi_b(x))\right]\Delta(x,y)=\delta^{(d)}(x-y)\nonumber
	\\& V''[\phi_b(x)]\approx V''[\phi_b(0)]+\ldots
	\\& \implies \left[-\partial^2+V''[\phi_b(0)]\right]\Delta(x,x+y)=\delta^{(d)}(y)+\ldots\nonumber
\end{align}
Thus to leading order this propagator is the same as for the effective-potential case \cite{PhysRevD.46.1671}. We find
\begin{align}
	-\frac{1}{3!\times 2!}\lambda^2\int d^3x d^3y \Delta(x,y)^3 \phi_b(x)\phi_b(y)\approx  	-\frac{1}{3!\times 2!}\lambda^2 \int d^3x \phi_b(x)^2 \int_{\abs{y}<R} d^3y \Delta(y)+\ldots\nonumber
\end{align}
The integral with respect to $y$ is the same as in Equation \ref{eq:EffPotInt}, and gives
\begin{align}
	\frac{d }{d \log \mu_3}S_\text{2-Loop}=-\frac{1}{96\pi^2\times 2!}\lambda^2 \int d^3x \phi_b(x)^2.
\end{align}
We see that the effective action is renormalization-scale invariant.

\begin{figure}[h!]
	\includegraphics[scale=0.8]{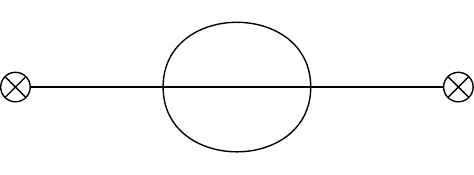}
	\caption{Two-loop diagram with external-current insertions.\label{fig:RealScalEffAc2LoopDyn}}
\end{figure}

Next, consider the dynamical prefactor. Up to a factor of $2 \pi$, the renormalization-scale dependence is
\begin{align}
	\frac{d}{d\log \mu_3}\sqrt{\abs{\kappa}}=-\frac{1}{2\sqrt{\abs{\kappa}}} \int d^3x \bU(x)^2 \frac{d}{d\log  \mu_3} m^2=- \frac{1}{96 \pi^2 \times 2!\sqrt{\abs{\kappa}}}\lambda^2\int d^3x \bU(x)^2,\nonumber
\end{align}
and the relevant 2-loop diagram is given in Figure \ref{fig:RealScalEffAc2LoopDyn}.

Using the Feynman rules from Section \ref{sec:FeynmanRules} this diagram is\footnote{The minus sign comes from the $\bU^n$ rule. Note also the sign of $\kappa$ as compared to $\abs{\kappa}$.}
\begin{align}
	-	\frac{1}{3!\times 2!}\kappa^{-1}\lambda^2\int d^3x d^3y \bU(x) \bU(y) \Delta(x,y)^3.
\end{align}

We find
\begin{align*}
-	\frac{d}{d\log \mu_3}\left\langle \lambda^2 \kappa^{-1}	\frac{1}{3!\times 2!}\int d^3x d^3y \bU(x) \bU(y) \Delta(x,y)^3\right\rangle =-\frac{1}{96\pi^2\times 2! \kappa}\lambda^2 \int d^3x \bU(x)^2.
\end{align*}

The dynamical prefactor is manifestly renormalization-scale invariant once we multiply the above diagram with the leading-order prefactor $\frac{\sqrt{\abs{\kappa}}}{2\pi}$.

Although this section only treated the real-scalar model, everything carries through once vector-bosons are included. The same methods can also be used to check that the Sphaleron rate is renormalization-scale invariant.

\subsection{Radiative barriers and the dynamical prefactor}

The previous section dealt with scale-dependence arising from two-loop diagrams. Consider now instead a model where a heavy field is integrated out. For example a vector boson. The procedure of integrating out particles, and consistently calculating the rate, has been studied in~\cite{Ekstedt:2021kyx,Gould:2021ccf,PhysRevD.46.1671,Andreassen:2016cvx,Weinberg:1992ds}.

This scenario is relevant as even if a barrier is absent at tree-level, it can be generated from loops~\cite{Arnold:1992rz,PhysRevD.7.1888}. As an example, consider a $\mathrm{SU}(2)$ gauge theory with a scalar in the fundamental representation; the leading-order potential is
\begin{align}
	V(\phi)=\frac{1}{2}m^2\phi^2-\frac{1}{16 \pi} e^3 \phi^3+\frac{1}{4} \lambda\phi^4.
\end{align}
The next order includes a scale-dependent contribution \cite{Ekstedt:2021kyx}
\begin{align}
	V_\text{NLO}\supset \frac{1}{1024 \pi^2}e^4\int d^3x\phi^2(x) \left\lbrace 51 \log\left(\frac{\mu_3^2}{ \phi^2(x)e^2}\right)-126 \log(3/2)+33\right\rbrace.
\end{align}
Normally the scale dependence in $V_\text{NLO}$ cancels with the beta function of the mass parameter:
\begin{align}
	\frac{d}{d \log \mu_3}m^2=-\frac{51}{256\pi^2}e^4+\ldots
\end{align}
This is manifestly the case for the statistical prefactor,  but it is more complicated for the dynamical prefactor. Indeed, naively the negative eigenvalue satisfies
\begin{align}
	\left[-\partial^2+V''[\phi_b]\right]\bU=\kappa \bU,
\end{align}
which implies
\begin{align}
	\delta \kappa =\int  d^3x V''_\text{NLO}[\phi_b(x)] \bU(x)^2.
\end{align}

Let us check that this procedure is correct by using the methods in Section \ref{sec:DynamicalPrefac}. Treating $V_\text{NLO}$ as a perturbation, the scale-dependent piece in $\zeta_1$ is (see Equation \ref{eq:Zeta1})
\begin{align}
	\delta \zeta_1=(2\kappa)^{-1}\frac{51}{256\pi^2}e^4 u \log\mu_3.
\end{align}
The corresponding change to the rate is
\begin{align}
	\delta\Gamma=\frac{\sqrt{ \abs{\kappa}}}{\sqrt{2\pi}} (2\kappa)^{-1}\left[\frac{51}{256\pi^2}e^4 \log \mu_3\right]e^{-\bar{S}_\text{eff}}.
\end{align}
As shown above, the scale-dependent piece in $\kappa$ is $\frac{d}{d \log \mu_3}\kappa=\frac{51}{256 \pi^2}e^4$.
This implies
\begin{align}
	\delta A_\text{dyn}=-\frac{\sqrt{ \abs{\kappa}}}{\sqrt{2\pi}} (2\kappa)^{-1}\frac{51}{256\pi^2}e^4 .
\end{align}

Using the contribution from $\delta \zeta_1$, we see that the dynamical prefactor is renormalization-scale invariant.

\subsection{Factorization of the rate}\label{sec:FinalFacRate}
In the previous sections we have seen that the rate factorizes as
\begin{align}
	\Gamma=A_\text{dyn} \times A_\text{stat}.
\end{align}

The statistical prefactor is
\begin{align}\label{eq:FinalStatistical}
	A_\text{stat}= \int_{u=0} \mathcal{D} \phi e^{-S[\phi+\phi_b]}\times J_{ZM}\equiv e^{-\overline{S}_\text{eff}[\phi_b]}.
\end{align}
Here $\overline{S}_\text{eff}[\phi_b]$ omits all contributions from negative and zero eigenmodes. The zero modes also generate operator insertions; these are included in $J_{ZM}$ as discussed in Section \ref{sec:ZeroModes}. For concrete calculations, the Feynman rules given in Section \ref{sec:FeynmanRules} can be used.

The rate should be normalized with $e^{-\int d^3x V_\text{eff}(\phi_S)}$, where $V_\text{eff}(\phi_S)$ denotes the effective potential in the metastable state.\footnote{This last point is important since $A_\text{stat}$ is not finite without the normalization. In practice $V_\text{eff}(\phi_S)$ should be included diagram-by-diagram as discussed in Section \ref{sec:PracStat}.}

The dynamical prefactor is defined as
\begin{align}\label{eq:FinalDynamical}
 A_\text{dyn}=\frac{1}{\eta \times A_\text{stat}}\int_{u=0} \mathcal{D}\phi  \left[\int d^3x \bU(x) \delta_{\phi(x)}\zeta(\phi)\right]e^{-S[\phi+\phi_b]}\times J_{ZM} .
\end{align}

The virtue of these definitions is that the renormalization-scale dependence for $A_\text{dyn}$ and $A_\text{stat}$ are simple; both $A_\text{stat}$ and $A_\text{dyn}$ are manifestly real to all orders; $A_\text{dyn}$ and $A_\text{stat}$ can be calculated independently; there is a clear physical interpretation with  $A_\text{stat}$ as the Boltzmann suppression, and $A_\text{dyn}$ as the rate of probability-flow across the saddle-point.

To leading order $A_\text{dyn}$ can be interpreted as the growth-rate of the bubble.

\section{Considerations for concrete calculations}\label{sec:Practical}
This section shows how to calculate propagators in an inhomogeneous bounce background; what diagrams appear at two loops; and how to isolate divergences in dimensional regularization.

\subsection{Propagators}\label{sec:PracticalGreen}
The propagator satisfies the equation
\begin{align}\label{eq:GreenOriginalDef}
	\left[-\partial^2+V''[\phi_b(x)]\right]\Delta(x,y)=\delta^{3}(x-y).
\end{align}
Because the bounce is spherically symmetric, it is useful to expand the propagator in spherical harmonics\cite{1993}
\begin{align}\label{eq:GreenSpherical}
	\Delta(x,y)=\frac{1}{4\pi } \sum_{l=0}^\infty (2l+1) P_l(\cos\alpha)\Delta_l (r,r'),
\end{align}
with the short-hand $r\equiv \abs{\vec{x}},~r'\equiv \abs{\vec{y}}$ and $\cos\alpha =\frac{\vec{x}\cdot \vec{y}}{\abs{\vec{x}}\abs{\vec{y}}}$.

Here $\Delta_l (r,r')$ satisfies
\begin{align}\label{eq:GreensHarmonic}
	\left[-\partial_r^2-\frac{2}{r}\partial_r +\frac{l(l+1)}{r^2}+V''[\phi_b(r)]\right]\Delta_l(r,r')=-\delta(r-r')/r^2.
\end{align}
Given the bounce, Equation \ref{eq:GreensHarmonic} can be solved numerically for each $l$.

There are two caveats. First, for practical reasons one can not solve Equation \ref{eq:GreensHarmonic} for infinitely many $l$. Also, the propagator diverges in the $\vec{x}\rightarrow \vec{y}$ limit. 

Both of these problems have a common solution\cite{Dunne:2005te,Dunne:2005rt}. The idea is to solve Equation \ref{eq:GreensHarmonic} for $l\in (1,L)$ where $L$ is typically between 50 and 100. The remaining part of the sum can then be done analytically by using the WKB approximation.

There are two cases. First, assume that $\cos\alpha <1$. Solving Equation \ref{eq:GreensHarmonic} (see Appendix \ref{app:GreensFunctions}) for large $l$ one finds to leading order
\begin{align}\label{eq:GreenDiffAngle}
	&\frac{1}{4\pi } \sum_{l=L+1}^\infty (2l+1) P_l(\cos\alpha)\Delta_l (r,r')=\delta_L+\frac{1}{4\pi }\frac{1}{\abs{\vec{x}-\vec{y}}},
	\\& \delta_L=-\frac{1}{4\pi}\sum_{l=0}^L (2l+1)\Delta_l(r,r') P_l(\cos\alpha).
\end{align}
The $\delta_l$ term can be evaluated numerically for given $L$.

For the second case we assume $\cos\alpha=1$, and find
\begin{align}\label{eq:GreenSameAngle}
	&\frac{1}{4\pi } \sum_{l=L+1}^\infty (2l+1) P_l(1)\Delta_l (r,r')=\delta_L+\frac{1}{4\pi}\frac{\Gamma(1-2\epsilon)}{\Gamma(1-\epsilon)}\left(\mu^2 e^{\gamma}\right)^{\epsilon}(\abs{\abs{x}-\abs{y}})^{2\epsilon-1}
	\\& \delta_L=
	\begin{cases}
		-\frac{1-\left(\abs{y}/\abs{x}\right)^{L+1}}{4 \pi (\abs{x}-\abs{y})}       & \quad \text{if } \abs{x}>\abs{y}\\
		-\frac{1-\left(\abs{x}/\abs{y}\right)^{L+1}}{4 \pi (\abs{y}-\abs{x})} & \quad \text{if }\abs{x}< \abs{y}
	\end{cases}
\end{align}
Note that the leading $\delta_L$ term is finite in the $\abs{x} \rightarrow \abs{y}$ limit. Higher-order terms are given in Appendix \ref{app:GreensFunctions}.

Zero modes only contribute when $l=1$. To remove the zero-modes, start with Equation \ref{eq:GreenOriginalDef}
\begin{align}
	\left[-\partial^2+V''[\phi_b(x)]\right]\Delta(x,y)=\delta^{3}(x-y).
\end{align}
Now use the representation
\begin{align}
	\Delta(x,y)=\sum_a \frac{\bV_a(x)\bV_a(y)}{\lambda_a},
\end{align}
where $\bV_a(x)$ is an eigenfunction of $	\left[-\partial^2+V''[\phi_b(x)]\right]$ with eigenvalue $\lambda_a$. From Section \ref{sec:ZeroModes} we know that the normalized zero-modes are\cite{Andreassen:2016cvx,Andreassen:2017rzq}
\begin{align}
	\bV_\mu=	(S_B)^{-1/2} \partial_\mu \phi_b.
\end{align}
This leads us to consider the equation
\begin{align}
	\left[-\partial^2+V''[\phi_b(x)]+\epsilon\right]\Delta^\epsilon(x,y)=\delta^{3}(x-y),
\end{align}
where we can remove the zero eigenvalues by defining 
\begin{align}
	\Delta'(x,y)=\lim_{\epsilon\rightarrow 0}\left[\Delta^\epsilon(x,y)-(S_B)^{-1}\sum_\mu \frac{\partial_\mu \phi_b(x) \partial_\mu \phi_b(y)}{\epsilon}\right].
\end{align}

This modified propagator, also known as a generalized Green's function, satisfies
\begin{align}
	\left[-\partial^2+V''[\phi_b(x)]\right]\Delta'(x,y)=\delta^{3}(x-y)-(S_B)^{-1} \partial_\mu \phi_b(x) \partial_\mu \phi_b(y) .
\end{align}

Again, this equation can be solved by expanding $\Delta'(x,y)$ in spherical harmonics; as mentioned, the $ \partial_\mu \phi_b(x) \partial_\mu \phi_b(y)$ term only contributes to $l=1$. 

In summary:
\vspace*{-0.25cm}
\begin{enumerate}
	\itemsep0.5em 
	\item Solve the bounce equation $	\partial^2 \phi_b(x)=V'(\phi_b)$ numerically
	\item Use Equations \ref{eq:GreenSpherical} and \ref{eq:GreensHarmonic} to find the propagator for $l=0,\ldots L$. Choose $L$ according to the required precision.
	\item Introduce an angular cut-off $\delta \ll 1$, and define $\cos\alpha=\frac{\vec{x}\cdot \vec{y}}{\abs{\vec{x}}\abs{\vec{y}}}$.
	\item If $\alpha \geq \delta$, use Equation \ref{eq:GreenDiffAngle} to do the sum from $l=L+1$ to $\infty$.
	\item If $\alpha < \delta$, use Equation \ref{eq:GreenSameAngle} to do the sum from $l=L+1$ to $\infty$.
	\item When calculating integrals of the form $\int d^3x d^3y \Delta(x,y)$, use the result from point 5 to analytically calculate the contribution from the $\vec{x}\rightarrow \vec{y}$ region. 
	\item All $\epsilon$ poles, with $d=3-2\epsilon$, come from the $\vec{x}\rightarrow \vec{y}$ region.
	\item Integrate over the remaining phase-space numerically.
\end{enumerate}

Finally, external-current insertions can be omitted after replacing
\begin{align}
	\Delta(x,y)\rightarrow \Delta(x,y)-\frac{\bU(x)\bU(y)}{\kappa}.
\end{align}

\subsection{The statistical prefactor}\label{sec:PracStat}
Consider now possible diagrams appearing at the two-loop level. As before we use the real-scalar model defined in Section \ref{sec:FieldTheory} with a large damping.

Diagrams contributing to the statistical prefactor can be divided into three classes: vacuum diagrams, external-current insertions, and zero-mode insertions. Note that at NLO\te equivalent to two loops for equilibrium observables\te there are contributions from lower-loop diagrams with operator-insertions.

The Feynman rules are the same as in Section \ref{sec:FeynmanRules}, and $\kappa$ denotes the negative eigenvalue, while $\bU(x)$ is the corresponding eigenvector normalized as $\int d^3x \bU(x)^2=1$. All propagators are defined with zero-modes removed. 

\subsubsection{Vacuum diagrams}
There are three vacuum diagrams at two loops, which are given in Figure \ref{fig:RealScalEffAc2Loop}. The bubble and sunset diagrams are only finite after normalizing with the effective potential evaluated in the metastable state. As such, for actual calculations one should always add the corresponding diagram for the effective potential with a relative minus sign.  For example, the double-bubble diagram in Figure \ref{fig:RealScalEffAc2Loop} gives
\begin{align}
	\frac{1}{8}\lambda \int d^3x \Delta(x,x)^2.
\end{align}
The divergence comes from the $\abs{x}\rightarrow \infty$ region. For numerical stability one should include the effective-potential term directly:
\begin{align}
	\frac{1}{8}\lambda \int d^3x \Delta(x,x)^2\rightarrow \frac{1}{8}\lambda \int d^3x \left[\Delta(x,x)^2-\Delta_S(x,x)^2\right],
\end{align}
where $\Delta_S(x,x)$ is the propagator for a constant $\phi=\phi_S$.

Since all diagrams exponentiate we can directly include them in the effective action. This gives an extra minus sign for each diagram. To estimate the size of each diagram we use the power-counting
\begin{align}
	\Delta(x,y)\sim m, \quad \phi_b^2\sim m^2\lambda^{-1},\quad d^3 x\sim m^{-3} ,\quad m^2 \sim \lambda\sim g^2, \quad \frac{m}{\lambda}\gg 1.
\end{align}
The leading-order action then scales as $\SLO\sim \frac{m}{\lambda}$. At NLO the relevant contributions scale as $\frac{\lambda}{m}$.

The double-bubble diagram in Figure \ref{fig:RealScalEffAc2Loop} gives
\begin{align}
	\frac{1}{8}\lambda \int d^3x \Delta(x,x)^2\sim \frac{\lambda}{m},
\end{align}
and the dumbbell diagram gives
\begin{align}
	-\frac{1}{8} \int d^3 x d^3 y (\lambda \phi_b(x)-g)(\lambda \phi_b(y)-g) \Delta(x,x) \Delta(x,y)\Delta(y,y)\sim \frac{\lambda}{m}.
\end{align}
Finally, the sunset diagram scales as
\begin{align}
	-\frac{1}{3!\times 2!} \int d^3 x d^3 y (\lambda \phi_b(x)-g)(\lambda \phi_b(y)-g) \Delta(x,y)^3\sim \frac{\lambda}{m}.
\end{align}

\subsubsection{External-current insertions}
\begin{figure}
\begingroup
\begin{alignat*}{3}
	&\begin{gathered}
		\includegraphics[width=0.2\textwidth]{Bubble2Ext}
	\end{gathered}
	\quad
	&& \begin{gathered}
		\includegraphics[width=0.25\textwidth]{4ExternalCurrents}
	\end{gathered}
	\quad
	&&\begin{gathered}
		\includegraphics[width=0.3\textwidth]{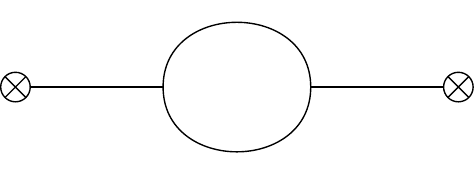}
	\end{gathered}
	\\
	&\begin{gathered}
		\includegraphics[width=0.2\textwidth]{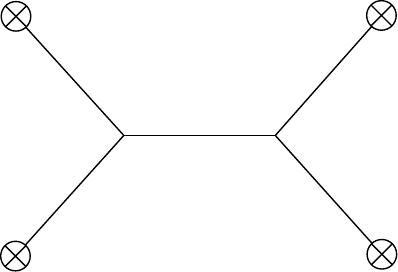}
	\end{gathered}
	\quad
	&& \begin{gathered}
		\includegraphics[width=0.2\textwidth]{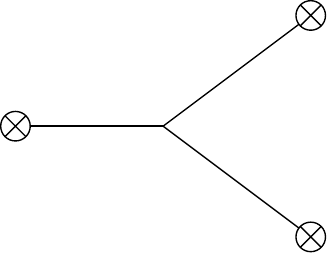}
	\end{gathered}
	\quad
	&&\begin{gathered}
		\includegraphics[width=0.2\textwidth]{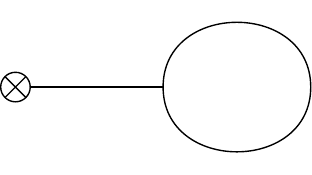}
	\end{gathered}
	\\
\end{alignat*}
\endgroup
	\caption{NLO diagrams contributing to the statistical prefactor.\label{fig:RealScalEffAc2LoopStat}}
\end{figure}

Here diagrams are labeled by the order they appear, left to right and top to bottom. All diagrams scale as $\frac{\lambda}{m}$. The diagrams are given in Figure \ref{fig:RealScalEffAc2LoopStat}.
\begin{alignat*}{2}
	&D_1=-\frac{1}{4}\lambda \kappa^{-1}\int d^3x \Delta(x,x)\bU(x)^2, \quad && D_2=3\frac{\lambda}{4!} \kappa^{-2}\int d^3x \bU(x)^4,
	\\&
	D_3= \frac{1}{4}\kappa^{-1}\int d^3x d^3y A(x,y)\Delta(x,y)^2 \bU(x) \bU(y), \quad && D_4=-\frac{3}{4!}\kappa^{-2}\int d^3x d^3y A(x,y)\Delta(x,y)\bU(x)^2\bU(y)^2,
	\\& D_5=\frac{15}{72} \kappa^{-3}\left(\int d^3x (\lambda\phi_b(x)-g) \bU(x)^3\right)^2, \quad &&D_6=\frac{1}{8}\kappa^{-1}\left(\int d^3x (\lambda\phi_b(x)-g)\Delta(x,x) \bU(x)\right)^2,
\end{alignat*}
\begin{align*}
 D_7=\frac{3}{6} \kappa^{-2}\left(\int d^3x (\lambda\phi_b(x)-g) \bU(x)^3\right)\left(\int d^3x (\lambda\phi_b(x)-g)\Delta(x,x) \bU(x)\right) .
\end{align*}

In addition, the notation $A(x,y)\equiv (\lambda\phi_b(x)-g)(\lambda\phi_b(y)-g)$ is used. In the above, $D_5$ represents the square of diagram 5, $D_6$ the square of diagram 6, and $D_7$ the product of diagrams 5 and 6. Note that all diagrams are finite in the $\abs{\vec{x}}\rightarrow \infty$ limit.

\subsubsection{Zero-mode insertions}
As noted in Section \ref{sec:ZeroModes}, when removing zero modes we have to include the operators given in Equation \ref{eq:ZeroModeOperator}. We denote these operators with black circles to differentiate them from other operators. In addition, external-current insertions are omitted for brevity.

\begin{figure}[h]
	\begingroup
	\begin{alignat*}{3}
		&\begin{gathered}
			\includegraphics[width=0.25\textwidth]{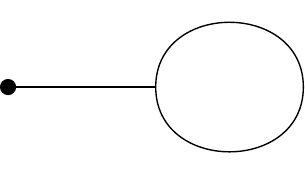}
		\end{gathered}
		\quad
		&& \begin{gathered}
			\includegraphics[width=0.2\textwidth]{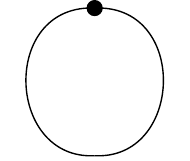}
		\end{gathered}
		\quad
		&&\begin{gathered}
			\includegraphics[width=0.2\textwidth]{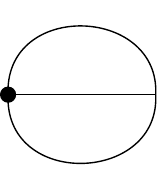}
		\end{gathered}
	\end{alignat*}
	\endgroup
	\caption{Zero-mode diagrams contributing to the statistical prefactor.\label{fig:RealScalEffAc2LoopZM}}
\end{figure}

The relevant diagrams are shown in Figure \ref{fig:RealScalEffAc2LoopZM}.
In terms of propagators these diagrams give
\begin{align}
	&D_1=\frac{1}{2}(\SLO)^{-1}\int d^3x  d^3y \partial^2\phi_b(x) \Delta(x,y)\Delta(y,y),\nonumber
	\\& D_2=\frac{1}{2}(\SLO)^{-2} \int d^3x d^3y \Delta(x,y)\left[\partial^2\phi_b(x)\partial^2\phi_b(y)-\partial_\mu \partial_\nu \phi_b(x)\partial_\mu \partial_\nu \phi_b(y)\right],
	\\& D_3=\frac{1}{3!}(\SLO)^{-3}\int d^3x d^3y d^3z d^3w (\lambda \phi_b(x)-g)\Delta(x,w)\Delta(y,w)\Delta(z,w) \overline{A}(x,y,z),\nonumber
\end{align}
where
\begin{align}
	\overline{A}(x,y,z)=&2 \partial_\mu \partial_\nu \phi_b(x) \partial_\nu \partial_\alpha \phi_b(y) \partial_\alpha \partial_\mu \phi_b(z)+\partial^2 \phi_b(x)\partial^2 \phi_b(y)\partial^2 \phi_b(z)\nonumber
	\\& -\partial^2 \phi_b(x)\partial_\mu \partial_\nu \phi_b(y)\partial_\mu \partial_\nu \phi_b(z)-\partial^2 \phi_b(y)\partial_\mu \partial_\nu \phi_b(x)\partial_\mu \partial_\nu \phi_b(z)
	\\&-\partial^2 \phi_b(z)\partial_\mu \partial_\nu \phi_b(x)\partial_\mu \partial_\nu \phi_b(y).\nonumber
\end{align}

Note that $D_1 \sim D_2 \sim \frac{\lambda}{m}$, while $D_3 \sim \frac{\lambda^2}{m^2}$. This means that only $D_1$ and $D_2$ are relevant at NLO.

\subsection{Dynamical prefactor}
The dynamical prefactor is given in Section \ref{sec:DynamicalPrefac}. There are two types of contributions. Those coming from a quartic vertex and those coming from two cubic vertices.

The quartic contribution is given in Equation \ref{eq:QuarticDynamical} and gives
\begin{align}
	\delta A_\text{dyn}=\frac{\lambda \sqrt{\ak}}{2\pi \eta}\left[\frac{3}{8 \kappa^2}\int d^3x \bU(x)^4-\frac{1}{4\kappa} \int d^3x \Delta(x,x) \bU(x)^2\right].
\end{align}

\begin{figure}

\begingroup
\begin{alignat*}{3}
	&\begin{gathered}
		\includegraphics[width=0.2\textwidth]{TadpoleDyn1}
	\end{gathered}
	\quad
	&& \begin{gathered}
		\includegraphics[width=0.2\textwidth]{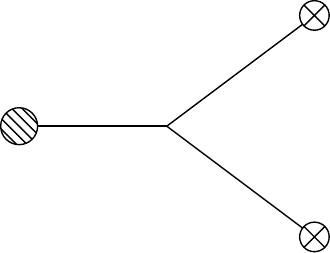}
	\end{gathered}
\end{alignat*}
\endgroup
	\caption{Diagrams contributing to the NLO dynamical prefactor.\label{fig:RealScalNLODyn}}
\end{figure}

The double-cubic contribution has one part coming from Equation \ref{eq:Ansatzxi2},  and another from \ref{eq:DynZeta2FT}. The contribution from Equation \ref{eq:Ansatzxi2} results in two diagrams, which are given in Figure \ref{fig:RealScalNLODyn}.
The first diagram is
\begin{align*}
	-\frac{1}{4 \kappa }\int d^3x d^3y\bU(x)A(x,y)\left[\Delta_{2\kappa}(x,y)-\Delta(x,y)\right]\Delta(y,y)+\frac{1}{2}\kappa^{-2}\int d^3x d^3y\bU(x)^3 \bU(y)\Delta(y,y)A(x,y),
\end{align*}
and the second is
\begin{align*}
\frac{1}{2 }\kappa^{-2}\int d^3x d^3y\bU(x)A(x,y)\left[\Delta_{2\kappa}(x,y)-\Delta(x,y)\right]\bU(y)^2-\kappa^{-3}\int d^3x d^3y\bU(x)^3 \bU(y)^3A(x,y).
\end{align*}

\subsubsection{Zero-mode insertions}
\begin{figure}
\begingroup
\begin{alignat*}{3}
	&\begin{gathered}
		\includegraphics[width=0.2\textwidth]{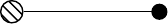}
	\end{gathered}
	\quad
	&& \begin{gathered}
		\includegraphics[width=0.2\textwidth]{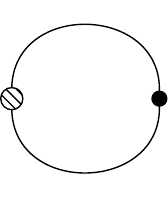}
	\end{gathered}
\end{alignat*}
\endgroup
	\caption{Zero-mode contributions for the dynamical prefactor.\label{fig:RealScalNLODynZM}}
\end{figure}

As mentioned, zero-mode operators also contribute to the dynamical prefactor. There are two one-loop diagrams (ignoring external-current insertions), these are given in Figure \ref{fig:RealScalNLODynZM}.

The first diagram gives
\begin{align*}
	-\kappa^{-2}(S_B)^{-1}\int d^3x \int d^3y\partial^2\phi_b(y)(\lambda \phi_b(x)-g) \bU(x)^2\left[\bU(y) \bU(x)-\frac{\kappa}{2}\left(\Delta_{2\kappa}(x,y)-\Delta(x,y)\right)\right]\sim \frac{\lambda}{m}.
\end{align*}
The second diagram scales as $\frac{\lambda^2}{m^2}$, and so is not relevant at this order.

\section{Discussion}

\subsection{Summary}
In this paper we have shown that the nucleation rate factorizes into an equilibrium and a non-equilibrium contribution\te a statistical and a dynamical prefactor. This factorization holds to all orders assuming an effective Langevin description. The formula derived in this paper lays the groundwork for calculating the nucleation rate beyond leading order. This is critical for getting a hold on uncertainties, and in particular, for estimating when the formalism breaks down.

To illustrate the calculations, this paper derived special Feynman rules. In addition, explicit calculations confirm that both the dynamical and the statistical prefactors are renormalization-scale invariant to two-loops.

\subsection{Future work}
This paper does not show that the dynamical prefactor is gauge invariant. Even though the statistical prefactor is gauge invariant~\cite{Lofgren:2021ogg,Hirvonen:2021zej,Garny:2012cg,Baacke:1999sc}, it is still important to verify that this holds for the full rate.

Another avenue is to apply the formalism of this paper to Sphaleron transitions. This is of great importance for Baryogenesis scenarios\cite{deVries:2017ncy}. In particular, the literature lacks a robust calculation of the Sphaleron rate with a radiative barrier. Moreover, current calculations of the functional determinant are limited to the Standard Model\cite{Baacke:1993aj,Carson:1989rf}. 

Lastly, this paper is based on classical field theory. This is motivated at high temperatures: integrating out high-energy modes results in a classical field theory with effective damping, couplings, and thermal noise. Explicit calculations of the damping are however sparse\cite{Bodeker:1999ey,Arnold:1998cy,Blaizot:1993zk,Moore:2000jw,Moore:1998swa,Gautier:2012vh}. There is also the question whether a Langevin description is applicable at higher orders. Answering these questions is essential for robust predictions.

\subsection{Conclusion}
It is the hope that the results of this paper will aid the theoretical understanding of phase transitions, and thereby reduce uncertainties for gravitational-wave predictions. Yet there are several open questions, and many problems remain. Solving these problems requires applying existing methods to new models; pushing high-temperature field theory to higher orders; and doing vital cross-checks with lattice computations.

\section*{Acknowledgement}
I would like to thank Oliver Gould, Joonas Hirvonen, and Johan Löfgren for discussions and a critical read-through of the manuscript. I am also indebted to Hamburg University and DESY for their hospitality. I am grateful to Aleksi Vuorinen and the Helsinki Institute of physics for their kind invitation to the University, where the final stages of this project took place. This work has been supported by the Swedish Research Council, project number VR:$2021$-$00363$. This work is supported by the Deutsche Forschungsgemeinschaft under Germany’s Excellence Strategy - EXC $2121$ Quantum Universe - $390833306$.

\appendix

\section{The dynamical prefactor for general damping}\label{app:GeneralDynamical}
Consider the potential
\begin{align}
	V(x+x_B)=V_B+\frac{1}{2}x_i \omega^{ij}x_j+\frac{1}{3!}\lambda^{ijk}_3x_i x_j x_k+\frac{1}{4!} \lambda_4^{ijkl}x_i x_j x_k x_l.
\end{align}
The goal is to solve Equation \ref{eq:ZetaEqGeneral} and find $\zeta$. That is, we need to solve
\begin{align}\label{eq:ZetaGeneralEq}
	\left[	\pi^i \partial_x^i+\left[\eta \pi ^i-\partial_x^i V\right]\partial_\pi^i-\eta T \partial_\pi^2\right] \zeta=0.
\end{align}
The leading-order solution, refereed henceforth as $\zeta_0$, is given in Equation \ref{eq:GeneralZeta}. 

Following Section \ref{sec:DynamicalPrefac}, make the ansatz $\zeta=\zeta_0+ \zeta_1 \zeta'_0(u)$. Plugging this ansatz into Equation \ref{eq:ZetaGeneralEq} one finds
\begin{align}\label{eq:Zeta1General}
		-T\eta \partial_\pi^2 \zeta_1+\left[\pi^i \partial_x^i+ \eta \pi^i \partial_\pi^i -x^i\omega^{ij}\partial_\pi^j+2\eta T \frac{u}{\gamma}\bU_\pi^i \partial_\pi^i\right]\zeta_1+\lambda \zeta_1=\bracket{\bU_\pi x^2}+\bracket{\bU_\pi x^3},
\end{align}
where $\bracket{x^3}\equiv \frac{1}{2!} \lambda_3^{ijk}x_i x_j x_k$, $\bracket{x^4}\equiv \frac{1}{3!} \lambda_4^{ijkl}x_i x_j x_k x_l$, and $\bU_\pi^i \bU_\pi^i=\frac{\lambda^2}{\kappa^2}$.

As in Section \ref{sec:DynamicalPrefac} the idea is to use a polynomial ansatz for $\zeta_1$, however, this ansatz is more involved than in the overdamped case. To see how things work, focus on the $\bracket{\bU_\pi x^3}$ term.

 The idea is to use the eigenbasis $\omega$: $\bV_a^i \omega_{ij}=\lambda_a \bV_a^j$. One can then expand $\pi^i=u_\pi \bU_x^i+\sum_{a\neq \kappa }\bV_a^i v_\pi^a$ and $x^i=u_x \bU_x^i+\sum_{a\neq \kappa }\bV_a^i v_x^a$.

The ansatz is
\begin{align}
	\zeta_1&=A+C_1 u_x^2+C_1^a u_x v_x^a+C_1^{ab} v_x^a v_x^b\nonumber
	\\& +E_1 u_\pi^2+E_1^a u_\pi v_\pi^a+E_1^{ab} v_\pi^a v_\pi^b
		\\& +F_1 u_\pi u_x+F_1^a u_x v_\pi^a+F_2^a u_\pi v_x^a+F_1^{ab} v_x^a v_\pi^b.\nonumber
\end{align}

Solving Equation \ref{eq:Zeta1General}, and keeping only terms that contribute to the rate, gives
\begin{align*}
&	E_1=\bracket{\bU_x^3}\frac{2 }{-6 \eta ^2 \kappa+27 \kappa^2}, \quad F_1=(-2 \beta -\lambda)E_1, \hspace{0.2cm}	C_1=\frac{1}{2} ((\beta +\lambda ) (2 \beta +\lambda )-2 \kappa ) E_1
\\&E_1^a=\bracket{\bU_x^2 \bV^a}\frac{\lambda}{ \kappa^2 (2\lambda+\eta)} \quad F_1^a=-\frac{2 \eta  \lambda +\kappa +3 \lambda ^2+\lambda_a}{\eta +2 \lambda }E_1^a
\\& F_2^a=\frac{-(\eta +\lambda ) (2 \eta +3 \lambda )+\kappa +\lambda_a}{\eta +2 \lambda } E_1^a, \quad C_1^a=\frac{\lambda_a (\eta +\lambda )+\lambda  (\eta +\lambda ) (2 \eta +3 \lambda )-\kappa  \lambda }{\eta +2 \lambda } E_1^a.
\end{align*}
Recall that $\lambda=\frac{1}{2}\left[\sqrt{\eta^2-4\kappa}-\eta\right]$. Other terms are given by similar formulas.
\subsection{Comparison with regular perturbation theory}

To check the results, consider $\omega^{ij}\rightarrow \omega^{ij}+\epsilon \omega_1^{ij}$ where $\epsilon$ is a small number. In this case we already know the answer because the correction to $\kappa$ follows from usual perturbation theory:
\begin{align}
	\delta \kappa =\bU_x^i \omega_1^{ij}  \bU_x^j,
\end{align}
where $\bU_x^i \omega_{ij}=\kappa \bU_x^j$.

 As before, make the ansatz $\zeta=\zeta_0 +\epsilon \zeta_1  \zeta'_0$ where $\zeta_0$ is given in Equation \ref{eq:GeneralZeta}.  After some algebra one finds
\begin{align}
	-T\eta \partial_\pi^2 \zeta_1+\left[\pi^i \partial_x^i+ \eta \pi^i \partial_\pi^i -x^i\omega^{ij}\partial_\pi^j+2\eta T \frac{u}{\gamma}\bU_\pi^i \partial_\pi^i\right]\zeta_1+\lambda \zeta_1=\bU_\pi^i \omega_1^{ij} x^j.
\end{align}

Following the previous section, the ansatz is
\begin{align}
	\zeta_1=A u +\sum_a A_a V^i_a \pi^i+B_1 \bU_x^i x^i+\sum_a B_a \bV^i_a  x^i v_a,
\end{align}
where $V^i_a$ are eigenvectors of $\omega^{ij}$ with positive eigenvalues $\lambda_a$. After collecting terms we get
\begin{align}
	&A=\frac{\bracket{\bU_\pi \bU_x}\ak}{2\lambda(\lambda^2+\ak)}, \quad B_1 =\frac{\bracket{\bU_\pi \bU_x}\lambda }{\lambda^2+\ak},
	\\& A_a=\frac{\bracket{\bU_\pi \bV^a}}{\lambda_a}, \quad B_a=\frac{\lambda+\eta}{\lambda_a} \bracket{\bU_\pi \bV^a}.\nonumber
\end{align}
We use the notation $\bracket{U V}\equiv U^i \omega_1^{ij} V^j$.

Just as in the overdamped case, the rate involves integrating over a surface normal to $u$. Since eigenvectors are orthogonal, all $A_a$ and $B_a$ terms drop out.

Adding $\zeta_1$ changes the rate by
\begin{align}
	\delta \Gamma=-\frac{1}{8\pi} \bracket{\bU_x^2} \left(\frac{4}{\sqrt{\eta ^2+4 \omega ^2}}-\frac{2}{\eta }\right)  \left\vert \det \omega \right\vert^{-1/2}e^{-V_B/T}
\end{align}

With $\zeta_1$ taken care of, this still leaves contributions from the Boltzmann factor.  This contribution is
\begin{align}
	\delta \Gamma=-\frac{\bracket{\bU_x^2}}{2\pi}\frac{ \left(\eta -\sqrt{\eta ^2+4 \omega ^2}\right)^2}{8 \eta  \omega ^2}  \left\vert \det \omega \right\vert^{-1/2}e^{-V_B/T}.
\end{align}

Adding these two contributions together is equivalent to taking $\ak \rightarrow \ak-\bracket{\bU_x^2}$ in
\begin{align}
	j=\frac{\lambda}{2\pi} \left\vert \det \omega \right\vert^{-1/2}e^{-V_B /T}.
\end{align}

This result agrees with regular perturbation theory.

\section{ The dynamical prefactor for dimension operators}\label{sec:Dim6}

Recall Equation \ref{eq:Zeta1}
\begin{align}
		T	\partial_i^2 \zeta_1+\partial_i \zeta_1 \left[2\kappa u \bU^i-\omega^{ij}x_j\right]+\kappa \zeta_1=\bracket{\bU x^2}+\bracket{\bU x^3},
\end{align}
where the ansatz was
\begin{align*}
		\zeta_1=A+ B_1 u + B_a v_a+C_1 u^2+ C_a u v_a +C_{ab}  v_{ab}+ D_1 u^3+ D_a u^2 v_a+ D_{ab} u v_{ab}+ D_{abc} v_{abc}.
\end{align*}
For brevity we use $v_{abc}\equiv v_a v_b v_c$ and $\bracket{\bV_{abc}}\equiv\bracket{ \bV_a \bV_b \bV_c}$; all repeated indices are summed over.

Identifying terms one finds ($\lambda_{ab\ldots}\equiv \lambda_a+\lambda_b+\ldots$)
\begin{alignat}{3}\label{eq:SolWithoutD6}
&D_{abc}=\bracket{\bU \bV_{abc}}(\kappa-\lambda_{abc})^{-1}, \quad && D_{ab}=3\bracket{\bU^2 \bV_{ab}}(2\kappa-\lambda_{ab})^{-1},\nonumber
\\&D_{a}=3\bracket{\bU^3 \bV_{a}}(3\kappa-\lambda_{a})^{-1} , \quad && D_{1}=\bracket{\bU^4 }(4\kappa)^{-1},\nonumber
\\& C_{ab}=\bracket{\bU \bV_{ab}}(\kappa-\lambda_{ab})^{-1},  \quad && C_{a}=2\bracket{\bU^2 \bV_{a}}(2\kappa-\lambda_{a})^{-1},
\\& C_{1}=\bracket{\bU^3 }(3\kappa)^{-1},  \quad && B_1=-T(2\kappa)^{-1}(6 D_1+ 2 D_{aa}),\nonumber
\\& B_a=-T \kappa^{-1}(6 D_{acc}),  \quad && A=-T \kappa^{-1}(2 C_{aa}+2 C_1).\nonumber
\end{alignat}

Consider now a potential
\begin{align}
	\bU^{i} \partial_i V=\bracket{\bU x^2}+\bracket{\bU x^3}+\bracket{\bU x^4}+\bracket{\bU x^5}.
\end{align}

In the language of SM-EFT the above potential would correspond to including a dimension 6 operator in the tree-level Lagrangian, see for example \cite{Croon:2020cgk}.

This new potential necessitates including the terms
\begin{align}\label{eq:SolWithD6}
	\delta \zeta_1&=E_1 u^4+ E_a u^3 v_a +E_{ab} u^2 v_{ab}+E_{abc} u v_{abc}+E_{abcd} v_{abcd}\nonumber
	\\&+F_1 u^5+ F_a u^4 v_a +F_{ab} u^3 v_{ab}+F_{abc} u^2 v_{abc}+u F_{abcd} v_{abcd}+ F_{abcde} v_{abcde}.
\end{align}

One finds
\begin{alignat}{2}
	&F_1=(6\kappa)^{-1}\bracket{\bU^6}, \quad && E_1=(5\kappa)^{-1}\bracket{\bU^5},\nonumber
\\& F_a=5(5\kappa-\lambda_a)^{-1}\bracket{\bU^5 \bV_a}, \quad &&  E_a=4(4\kappa-\lambda_a)^{-1}\bracket{\bU^4 \bV_a},\nonumber
\\& F_{ab}=10(4\kappa-\lambda_{ab})^{-1}\bracket{\bU^4 \bV_{ab}} ,\quad && E_{ab}=6(3\kappa-\lambda_{ab})^{-1}\bracket{\bU^3 \bV_{ab}},
\\& F_{abc}=10(3\kappa-\lambda_{abc})^{-1}\bracket{\bU^3 \bV_{abc}}, \quad && E_{abc}=4(2\kappa-\lambda_{abc})^{-1}\bracket{\bU^2 \bV_{abc}},\nonumber
\\& F_{abcd}=5(2\kappa-\lambda_{abcd})^{-1}\bracket{\bU^2 \bV_{abcd}}, \quad && E_{abcd}=(\kappa-\lambda_{abcd})^{-1}\bracket{\bU^1 \bV_{abcd}},\nonumber
\\& F_{abcde}=(\kappa-\lambda_{abcde})^{-1}\bracket{\bU^1 \bV_{abcde}}.\nonumber
\end{alignat}

Remaining terms are given by Equation \ref{eq:SolWithoutD6} with the modification
\begin{alignat}{2}
	&\delta D_{abc}=-5\times 4 F_{abcee}-2 F_{abc}, \quad && D_{ab}=-4\times 3 F_{abee}-3\times 2 F_{ab},\nonumber
	\\& D_{a}=-3\times 2F_{aee}-4\times 3F_{a}, \quad && D_{1}=-2 F_{ee}-5\times 4 F_{1},
	\\& \delta C_{ab}=-4\times 3 E_{abee}-2 E_{ab}, \quad && \delta C_{a}=-3\times 2E_{aee}-3\times 2E_{a},\nonumber
	\\& \delta C_{1}=-2 E_{ee}-4\times 3 E_{1}.\nonumber
\end{alignat}

The above factors, once included in the rate, can be written in terms of propagators. See Section \ref{sec:FieldTheory} for the details.
\vspace*{-0.8cm}
\section{Higher-order corrections to the dynamical prefactor}\label{sec:Zeta2}
Here we determine the $\zeta_2$ term in the ansatz\vspace*{-0.3cm}
\begin{align}
	\zeta=\zeta_0+\zeta_1 \zeta'_0+ \zeta_2 \zeta'_0+\ldots
\end{align} 

Assuming $\bU^{i} \partial_i V=\bracket{\bU x^2}+\bracket{\bU x^3}$, we get the equation
\begin{align}\label{eq:Zeta2}
T	\partial_i^2 \zeta_2+\partial_i \zeta_2 \left[2\kappa u \bU^i-\omega^{ij}x_j\right]+\kappa \zeta_2&=\left(\kappa \frac{u}{T}\zeta_1+\partial_u \zeta_1 \right)\left(\bracket{\bU x^2}+\bracket{\bU x^3}\right)\nonumber
\\& + \sum_a \frac{\partial}{\partial v_a}\zeta_1 \left(\bracket{\bV^a x^2}+\bracket{\bV^a x^3}\right),
\end{align}
where $\zeta_1$ is given in Equation \ref{eq:SolWithoutD6}. Note that in solving this equation, much work can be recycled from Section \ref{sec:Dim6}. The ansatz is\vspace*{-0.3cm}
\begin{align*}
	\zeta_2&=A+ B_1 u + B_a v_a+C_1 u^2+ C_a u v_a +C_{ab}  v_{ab}+ D_1 u^3+ D_a u^2 v_a+ D_{ab} u v_{ab}+ D_{abc} v_{abc}
	\\&E_1 u^4+ E_a u^3 v_a +E_{ab} u^2 v_{ab}+E_{abc} u v_{abc}+E_{abcd} v_{abcd}
	\\&+F_1 u^5+ F_a u^4 v_a +F_{ab} u^3 v_{ab}+F_{abc} u^2 v_{abc}+u F_{abcd} v_{abcd}+ F_{abcde} v_{abcde}
		\\&+ G_1 u^6+ G_a u^5 v_a +G_{ab} u^4 v_{ab}+G_{abc} u^3 v_{abc}+u^2 G_{abcd} v_{abcd}+ u G_{abcde} v_{abcde}.
	\\&+ H_1 u^7+ H_a u^6 v_a +H_{ab} u^5 v_{ab}+H_{abc} u^4 v_{abc}+u^3 H_{abcd} v_{abcd}+ u^2 H_{abcde} v_{abcde}+ u H_{abcdef} v_{abcdef}.
\end{align*}
Most of these terms were given in Appendix \ref{sec:Dim6}. Only the $G$ and $H$ terms are new. These are however readily found\footnote{All of these expressions should be symmetrised (completely symmetric) in their respective indices.}
\begin{align*}
&	H_1=(8\kappa)^{-1} \kappa^{-1} D_1 \bracket{\bU^4}
\\& H_a=(7\kappa-\lambda_a)^{-1} \kappa^{-1} \left[3 D_1 \bracket{\bU^3 \bV_a}+ D_a \bracket{\bU^4}\right]
\\& H_{ab}=(6\kappa-\lambda_{ab})^{-1} \kappa^{-1} \left[3 D_1 \bracket{\bU^2 \bV_{ab}}+3 D_a \bracket{\bU^3 \bV_a}+D_{ab} \bracket{\bU^4}\right]
\\& H_{abc}=(5\kappa-\lambda_{abc})^{-1} \kappa^{-1} \left[D_1 \bracket{\bU \bV_{abc}}+3 D_a \bracket{\bU^2 \bV_{bc}}+3 D_{ab} \bracket{\bU^3 \bV_c}+D_{abc} \bracket{\bU^4}\right]
\\& H_{abcd}=(4\kappa-\lambda_{abcd})^{-1} \kappa^{-1} \left[ D_a \bracket{\bU \bV_{bcd}}+3 D_{ab} \bracket{\bU^2 \bV_{cd}}+3D_{abc} \bracket{\bU^3 \bV_d}\right]
\\& H_{abcde}=(3\kappa-\lambda_{abcde})^{-1} \kappa^{-1} \left[  D_{ab} \bracket{\bU\bV_{cde}}+3D_{abc} \bracket{\bU^2 \bV_{de}}\right]
\\& H_{abcdef}=(2\kappa-\lambda_{abcdef})^{-1} \kappa^{-1} \left[  D_{abc} \bracket{\bU \bV_{def}}\right]
\end{align*}
\begin{align*}
	&G_1	=(7\kappa)^{-1} \kappa^{-1} \left[D_1 \bracket{\bU^3}+C_1 \bracket{\bU^4}\right],
	\\& G_a=(6\kappa-\lambda_a)^{-1} \kappa^{-1} \left[2 D_1 \bracket{\bU^2 \bV_a}+3 C_1 \bracket{\bU^3 \bV^a}+ D_a \bracket{\bU^3}+C_a \bracket{ \bU^4}\right],
	\\& G_{ab}=(5\kappa-\lambda_{ab})^{-1} \kappa^{-1}\left[ D_1 \bracket{\bU \bV_{ab}}+3 C_1 \bracket{\bU^2 \bV_{ab}}+ 2D_a \bracket{\bU^2 \bV_b}+3C_a \bracket{ \bU^3 \bV_b}+C_{ab}\bracket{\bU^4}+D_{ab}\bracket{\bU^3}\right],
	\\& \!\begin{aligned}[t] 
		G_{abc}=(4\kappa-\lambda_{abc})^{-1} \kappa^{-1} &\left[   C_1 \bracket{\bU \bV_{abc}}+ D_a \bracket{\bU \bV_{bc}}+3C_a \bracket{ \bU^2 \bV_{bc}}\right.
	\\& \left.+3 C_{ab}\bracket{\bU^3 \bV_c}+2 D_{ab}\bracket{\bU^2 \bV_c}+D_{abc}\bracket{\bU^3}\right],
\end{aligned}
	\\& G_{abcd}=(3\kappa-\lambda_{abcd})^{-1} \kappa^{-1}  \left[  C_a \bracket{ \bU \bV_{bcd}}+3 C_{ab}\bracket{\bU^2 \bV_{cd}}+ D_{ab}\bracket{\bU \bV_{cd}}+2D_{abc}\bracket{\bU^2 \bV_d}\right],
	\\& G_{abcde}=(2\kappa-\lambda_{abcde})^{-1} \kappa^{-1 }\left[   C_{ab}\bracket{\bU \bV_{cde}}+D_{abc}\bracket{\bU \bV_{de}}\right].
\end{align*}

Remaining terms are given via Equation \ref{eq:SolWithD6}. These expressions are long, so only terms relevant for the NLO dynamical prefactor are explicitly given. 

Terms that involve two cubic vertices are
\begin{align}\label{eq:NLOCubic}
&\!\begin{aligned}[t] 
		D_1=&-\frac{5}{18}\kappa^{-2}\bracket{\bU^3}^2-\frac{2}{3}\kappa^{-1} \frac{1}{\kappa-\lambda_{cc}}\bracket{\bU^3}\bracket{\bU \bV^{cc}}
		\\&-\frac{\lambda_c}{2\kappa}\frac{1}{(2\kappa-\lambda_c)^2}\bracket{\bU^2 \bV^c}\bracket{\bU^2 \bV^c}, \nonumber
\end{aligned}\\
&\!\begin{aligned}[t] 
D_{ab}=&-2\frac{1}{(\kappa-\lambda_{ab})(2\kappa-\lambda_{ab})}\bracket{\bU^3}\bracket{\bU \bV^{ab}}
\\&-2\frac{2\kappa+\lambda_{ab}}{(2\kappa-\lambda_a)(2\kappa-\lambda_b)(2\kappa-\lambda_{ab})}\bracket{\bU^2\bV^a}\bracket{\bU^2\bV^b}
\\&\quad 2\frac{1}{(2\kappa-\lambda_c)(2\kappa-\lambda_{ab})}\bracket{\bU^2\bV^c}\bracket{\bV^c \bV^{ab}}
\\& -2 \frac{1}{(\kappa-\lambda_{ab})(\kappa-\lambda_{cc})}\bracket{\bU \bV^{ab}}\bracket{\bU \bV^{cc}}
\\&-2\frac{\lambda_{ab}+\lambda_{cc}}{(2\kappa-\lambda_{ab}-\lambda_{cc})(2\kappa-\lambda_{ab})}\left(\frac{1}{\kappa-\lambda_{ac}}+\frac{1}{\kappa-\lambda_{bc}}\right)\bracket{\bU \bV^{ac}}\bracket{\bU \bV^{bc}},
\end{aligned}\\
&\!\begin{aligned}[t] 
	F_{abcd}=(3!)^{-1} \frac{\kappa}{T}\frac{1}{(2\kappa-\lambda_{abcd})}&\left[\frac{1}{\kappa-\lambda_{ab}}\bracket{\bU \bV^{ab}}\bracket{\bU \bV^{cd}}+\frac{1}{\kappa-\lambda_{cd}}\bracket{\bU \bV^{ab}}\bracket{\bU \bV^{cd}}\right. \nonumber
\\&\left. 2\frac{1}{\kappa-\lambda_{ac}}\bracket{\bU \bV^{ac}}\bracket{\bU \bV^{bd}}+\frac{1}{\kappa-\lambda_{ad}}\bracket{\bU \bV^{ad}}\bracket{\bU \bV^{bc}}\right. \nonumber
\\& \left.+ \frac{1}{\kappa-\lambda_{bc}}\bracket{\bU \bV^{ad}}\bracket{\bU \bV^{bc}}+\frac{1}{\kappa-\lambda_{bd}}\bracket{\bU \bV^{bd}}\bracket{\bU \bV^{ac}}\right],
\end{aligned}\\
&\!\begin{aligned}[t] 
	B_1=-T(2\kappa)^{-1}(6 D_1+2 D_{aa}).\nonumber
\end{aligned}
\end{align}

All formulas use the notation $\lambda_{ab\ldots}=\lambda_{a}+\lambda_{b}+\ldots$; $\bV^{ab\ldots}=\bV^{a}\bV^{b}\ldots$; and repeated indices are summed over \textit{only} once\te the sum does not include the negative eigenvalue $\kappa$.

To calculate the rate we need to perform Wick contractions of $F_{abcd}$, $D_{ab}$, and $B_1$. This gives
\begin{align}
\delta A_\text{dyn}=\frac{\sqrt{\ak}}{2 \pi \eta}\left[3	F_{aa bb}\lambda_a^{-1}\lambda_b^{-1}+D_{aa} \lambda_a^{-1}+B_1\right].
\end{align}

After simplifying one finds
\begin{align}\label{eq:DynamicalZeta2}
	\delta A_\text{dyn}=&\frac{\sqrt{\ak}}{2 \pi \eta}T\left[\frac{5}{6}\kappa^{-3}\bracket{\bU^3}^2-\kappa^{-2}\lambda_a^{-1}\bracket{\bU^3}\bracket{\bU \bV^{aa}}+\kappa^{-2}\left(\frac{2}{\lambda_a-2\kappa}-\frac{1}{2\lambda_a}\right)\bracket{\bU^2\bV^a}\bracket{\bU^2\bV^a}\right.\nonumber
	\\&\left. + \kappa^{-1}\lambda_a^{-1}(2\kappa-\lambda_c)^{-1}\bracket{\bU^2\bV^c}\bracket{\bV^c\bV^{aa}}-\kappa^{-1}\frac{\kappa+\lambda_{ab}}{\lambda_a \lambda_b (\lambda_{ab}-\kappa)}\bracket{\bU \bV^{ab}}\bracket{\bU \bV^{ab}}\right.\nonumber
	\\&\left.+\kappa^{-1}\frac{1}{2}\lambda_a^{-1}\lambda_b^{-1}\bracket{\bU \bV^{aa}}\bracket{\bU \bV^{cc}} \right].
\end{align}
Or written in terms of propagators
\begin{align*}
	\delta A_\text{dyn}=\frac{\sqrt{\ak}}{2 \pi \eta}T&\left[\frac{23}{6}\kappa^{-3}\bracket{\bU^3}^2-\kappa^{-2}\bracket{\bU}^3\bracket{\bU\Delta_{2\kappa}} -3\kappa^{-2}\bracket{\bU^3}\bracket{\bU\Delta}+3\kappa^{-2}\bracket{\bU^2\Delta_k \bU^2} \right.
	\\& \left.+\frac{3}{2}\kappa^{-2}\bracket{\bU^2\Delta \bU^2} +\kappa^{-1}\bracket{\bU \Delta}^2-\kappa^{-1}\bracket{\bU\Delta \Delta \bU}-\kappa^{-1}\bracket{\bU^2 \Delta_{2 \kappa} \Delta}\right].
\end{align*}
Here $\bracket{\bU^2\Delta \bU^2}\equiv \lambda_3^{ijk} \bU^i \bU^j \Delta^{kl}\lambda_3^{l n m}\bU^n \bU^m$ and $\bracket{\bU\Delta \Delta \bU}\equiv \lambda_3^{ijk} \bU^i \Delta^{jl}\Delta^{kn}\lambda_3^{l n m}\bU^m$. In this last step we have approximated $\frac{\kappa+\lambda_{ab}}{\lambda_{ab}-\kappa}\approx 1$. Note that this approximation fails if the bubble-wall is too thick.

For field theory the equivalent expression is
\begin{align}\label{eq:DynZeta2FT}
	\delta A_\text{dyn}=&\left[\frac{23}{24}\kappa^{-3}\left(\int d^3x \bU(x)^3 (\lambda\phi_b(x)-g)\right)^2-\frac{1}{4}\kappa^{-2}\int d^3x d^3y\bU(x)^3 \bU(y)\Delta_{2\kappa}(y,y)A(x,y) \right.\nonumber
	\\&\left.-\frac{5}{4}\kappa^{-2}\int d^3x d^3y\bU(x)^3 \bU(y)\Delta(y,y)A(x,y)\right.\nonumber
	\\&\left.\frac{3}{4}\kappa^{-2}\int d^3x d^3y \bU(x)^2\left(\Delta_{2\kappa}(x,y) 
	+\frac{1}{2}\Delta(x,y)\right)\bU(y)^2 A(x,y)\right.\nonumber
	\\& \left. \frac{1}{4}\kappa^{-1}\left(\int d^3x \bU(x)\Delta(x,x)(\lambda \phi_b(x)-g)\right)^2\right.\nonumber
	\\&\left.-\frac{1}{4}\int d^3x d^3y \bU(x)\Delta(x,y)^2\bU(y)A(x,y) \right.\nonumber
	\\&\left.-\frac{1}{4}\kappa^{-1}\int d^3x d^3y \bU(x)^2\Delta_{2\kappa}(x,y)\Delta(y,y)\right].
\end{align}

\section{WKB approximation for large $l$}\label{app:GreensFunctions}
In this section we approximate the propagator for large $l$. To that end, consider Equation \ref{eq:GreensHarmonic} with $r\neq r'$. Start by rewriting this equation with Langer's ansatz \cite{Langer:1937qr} $\Delta(r,r')=e^{a x}\Psi(r=e^x)$. Essentially this ansatz eliminates the $\partial_r$ term in Equation \ref{eq:GreensHarmonic} and makes a WKB ansatz possible. We work in $d=3-2\epsilon$ dimensions to regularize all $\vec{x}\rightarrow \vec{y}$ divergences. The equation for $\Psi(x)$ is\cite{Dunne:2005te,Dunne:2006ct,Dunne:2006ac}
\begin{align}
	&\Psi''(x)- A(x) \Psi(x)=0,\nonumber
	\\&A(x)=e^{2x} V''[\phi_b(e^x)]+\overline{l}^2,
	\\& \overline{l}=\frac{d+2l-2}{2}.\nonumber
\end{align}
Making a WKB ansatz\cite{Wentzel:1926aor,Kramers:1926njj,Langer:1937qr} one finds
\begin{align}
	& \Delta^{>}_l (r,r')=r^{-(d-2)/2}\frac{1}{\sqrt{\overline{l}}} r^{-\overline{l}} \left[A^{>}(r')e^{-S_0(r)}\right],
	\\& 	\Delta^{<}_l (r,r')=r^{-(d-2)/2}\frac{1}{\sqrt{\overline{l}}} r^{\overline{l}} \left[A^{<}(r')e^{S_0(r)}\right],
\end{align}
where $S_0[r]=\frac{1}{2 \overline{l}}\int_0^r dx x V''[\phi_b(x)]$. Here $\delta^{>}_l(r,r')$ is defined with $r>r'$ and $\delta^{<}_l(r,r')$ with $r<r'$.

Above we expanded in powers of $\overline{l}$, and used boundary conditions to ensure that $\Delta^{>}_l(r,r')$ is finite in the $r\rightarrow \infty$ limit, and that $\Delta^{<}_l(r,r')$ is finite in the $r\rightarrow 0$ limit.

The actual propagator is found by demanding
\begin{align}
	& \left[\Delta^{>}(r,r')-\Delta^{<}(r,r')\right]_{r=r'}=0,
	\\&\left[\partial_r \Delta^{>}(r,r')-\partial_r\Delta^{<}(r,r')\right]_{r=r'}=1/(r')^{d-1}.
\end{align}

Doing the matching one finds the propagator
\begin{align}\label{eq:ApproxGreen}
	\Delta_l (r,r')=
	\begin{cases}
		\frac{(r')^l r^{-d-l+2}}{d+2 l-2} -\frac{(r')^l}{r^{2-d-l}}\frac{(d+2l-2)\left[f(r)-f(r')\right]+r^2 W(r)+(r')^2 W(r')}{(d+2l-2)^3}      & \quad \text{if } r>r'\\
		\frac{r^l (r')^{-d-l+2}}{d+2 l-2}-\frac{(r)^l}{(r')^{2-d-l}}\frac{(d+2l-2)\left[f(r')-f(r)\right]+r^2 W(r)+(r')^2 W(r')}{(d+2l-2)^3}   & \quad \text{if }r< r'
	\end{cases}
\end{align}
Here $W(r)\equiv V''[\phi_b(r)]$ and $f(r)\equiv \int_0^r ds s W(s)$.

The $l=L+1,\ldots,\infty$ sum is now straightforward, and the leading result is given in Equation \ref{eq:GreenSameAngle}. The next-to-leading result is
\vspace*{-0.8cm}\begin{align*}
		\\& \delta_1=
	\begin{cases}
		\frac{2 f(\abs{x}) \log \left(\frac{-2 \sqrt{\abs{x}\abs{y}}+\abs{x}+\abs{y}}{\abs{x}-\abs{y}}\right)+4 f(\abs{y}) \tanh ^{-1}\left(\sqrt{\frac{\abs{y}}{\abs{x}}}\right)+\left(\text{Li}_2\left(\frac{\abs{y}}{\abs{x}}\right)-4 \text{Li}_2\left(\sqrt{\frac{\abs{y}}{\abs{x}}}\right)\right) \left(\abs{x}^2 W(\abs{x})+\abs{y}^2 W(\abs{y})\right)}{16 \pi  \sqrt{\abs{x} \abs{y}}}    & \quad \text{if } \abs{x}>\abs{y}\\
		\frac{2 f(\abs{y}) \log \left(\frac{-2 \sqrt{\abs{x} \abs{y}}+\abs{x}+\abs{y}}{\abs{y}-\abs{x}}\right)+4 f(\abs{x}) \tanh ^{-1}\left(\sqrt{\frac{\abs{x}}{\abs{y}}}\right)+\left(\text{Li}_2\left(\frac{\abs{x}}{\abs{y}}\right)-4 \text{Li}_2\left(\sqrt{\frac{\abs{x}}{\abs{y}}}\right)\right) \left(\abs{x}^2 W(\abs{x})+\abs{y}^2 W(\abs{y})\right)}{16 \pi  \sqrt{\abs{x} \abs{y}}} & \quad \text{if }\abs{x}< \abs{y}
	\end{cases}
\end{align*}